\documentclass[showpacs,prd,twocolumn,aps,nofootinbib,superscriptaddress]{revtex4-1}

\usepackage{amsmath}
\usepackage{amssymb}
\usepackage{latexsym}
\usepackage{amsfonts}
\usepackage{epsfig}
\usepackage{psfrag}
\usepackage{graphicx}
\usepackage{wasysym}
\usepackage{ulem}
\usepackage{color}

\newcommand{\Eqref}[1]{Eq.~\eqref{#1}}

\renewcommand{\vec}[1]{\mathbf{#1}}

\begin{document}

\title{Magnetically amplified light-shining-through-walls via virtual
  minicharged particles}

\author{Babette D\"obrich}
\altaffiliation{Now at DESY, Notkestra\ss e 85, D-22607 Hamburg, Germany.}
\author{Holger Gies}
\affiliation{Theoretisch-Physikalisches Institut,
Friedrich-Schiller-Universit\"at Jena, Max-Wien-Platz 1, D-07743 Jena, Germany}
\affiliation{Helmholtz-Institut Jena, Fr\"obelstieg 3, D-07743 Jena, Germany}
\author{Norman Neitz}
\altaffiliation{Now at Max-Planck-Institut f\"ur Kernphysik, Saupfercheckweg 1, D-69117 Heidelberg, Germany.}
\affiliation{Theoretisch-Physikalisches Institut,
Friedrich-Schiller-Universit\"at Jena, Max-Wien-Platz 1, D-07743 Jena, Germany}
\author{Felix Karbstein}
\affiliation{Theoretisch-Physikalisches Institut,
Friedrich-Schiller-Universit\"at Jena, Max-Wien-Platz 1, D-07743 Jena, Germany}
\affiliation{Helmholtz-Institut Jena, Fr\"obelstieg 3, D-07743 Jena, Germany}

\begin{abstract}
  We show that magnetic fields have the potential to significantly enhance a
  recently proposed light-shining-through-walls scenario in quantum-field theories with
  photons coupling to minicharged particles. Suggesting a dedicated laboratory
  experiment, we demonstrate that this particular tunneling scenario could provide access to
  a parameter regime competitive with the currently best direct laboratory limits on minicharged
  fermions below the $\mathrm{meV}$ regime. With present day technology, such
  an experiment has the potential to even overcome the best model-independent cosmological
  bounds on minicharged fermions with masses below $\mathcal{O} (10^{-4}) \mathrm{eV}$.
\end{abstract}

\date{\today}


\maketitle

\section{Introduction}

``Light-shining-through-wall'' (LSW) experiments have become a valuable 
tool in the search for numerous theoretically well-motivated `weakly interacting slim particles' (WISPs) that could exist beyond the standard model of particle physics.
Here, the shorthand `WISP' subsumes hypothetical light particles at the (sub-)$\mathrm{eV}$ scale which exhibit a weak effective coupling to the 
electromagnetic field and are thus sensitive to optical searches \cite{Gies:2007ua,Jaeckel:2010ni}. Such particles generically arise in a number of extensions of the standard model
and under certain conditions also constitute a viable dark matter candidate \cite{Cheung:2007ut,Feldman:2007wj,Arias:2012mb,Cline:2012is}. 
Most prominently, LSW experiments provide for the best \textit{direct} 
laboratory bounds on axions, axion-like particles (ALPs) \cite{Chou:2007zzc,Robilliard:2007bq,Fouche:2008jk,Afanasev:2008jt,Pugnat:2007nu,Battesti:2010dm,Ehret:2010mh}
and massive hidden-sector photons (also referred
to as paraphotons) \cite{Okun:1982xi}, as well as for \textit{indirect} limits
on massless hidden-sector photons and fractionally 
charged or `minicharged' particles \cite{Holdom:1985ag,Anselm:1986gz,Gies:2006ca}.
See, e.g., \cite{Redondo:2010dp} and references 
therein for a recent overview.

\begin{figure}
\includegraphics[width=0.3\textwidth]{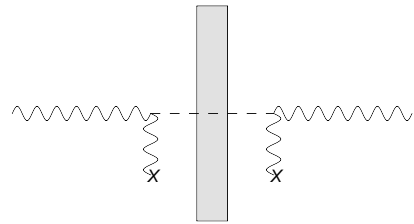} 
\caption{Classic LSW scenario with ALPs. In the presence of an external magnetic field (represented by wiggly lines ending at crosses) a 
photon entering from the left hand side
 (wiggly line) can be converted into an ALP (dashed line).
Due to its weak coupling to matter, the ALP can traverse a light-blocking barrier nearly unhindered and can subsequently be reconverted into a photon.}
\label{fig:lsw_WISP}
\end{figure}

In classic LSW scenarios the barrier is `tunneled' by {\it real}, i.e., on-shell WISPs. Their paradigm is the LSW scenario with ALPs
\cite{Sikivie:1983ip,VanBibber:1987rq}. 
Light-shining-through-walls becomes possible if the probe photons are converted into ALPs in front of the wall
and are reconverted into photons behind the wall. In typical laboratory searches
the conversion processes are induced by strong dipole magnets, see, e.g., \cite{Ehret:2010mh,Afanasev:2008jt,Steffen:2009sc,Robilliard:2007bq,Pugnat:2007nu}.
We depict this scenario in Fig.~\ref{fig:lsw_WISP}.

Besides LSW experiments, polarimetric measurements \cite{Maiani:1986md,Raffelt:1987im} carried out in various setups
\cite{Zavattini:2007ee,Cameron:1993mr,battesti} constitute an important means in the general search for WISPs.
Recent suggestions involve, e.g., advanced interferometric setups \cite{Guendelman:2009kv,Zavattini:2008cr,Dobrich:2009kd},
tests of Coulomb's law \cite{Jaeckel:2009dh,Jaeckel:2010xx}
or experiments utilizing high-intensity lasers \cite{Tommasini:2009nh,Dobrich:2010hi,Homma:2011pg}. 

In addition, also LSW experiments using astrophysical sources,
particularly the sun, have been carried out with remarkable success
as, e.g., in the CAST experiment \cite{Arik:2008mq}, see also
\cite{Irastorza:2011gs}. Such LSW `helioscopes' as well as arguments
involving stellar energy loss observations, lead to extremely strong
constraints \cite{Davidson:2000hf,Ahlers:2009ru,Burrage:2009yz}.  The
drawback is that most of these constraints are somewhat model
dependent, as their derivation involves extrapolations to a different
momentum-transfer regime \cite{Masso:2006gc,Jaeckel:2006xm}. Further
{\it indirect} bounds on various hypothetical particles can be derived
from cosmology by studying possible distorsions of the cosmic
microwave background (CMB) \cite{Melchiorri:2007sq}. The 
model-independent limits of these bounds for minicharged particles are
currently somewhat stronger than the direct laboratory bounds.

The high sensitivity of laboratory LSW experiments stems from
the large number of laser photons available for the conversion 
process in front of the wall, facing the possibility of even single photon detection on its rear side.
Currently, the best experimental bounds are due to the first stage of the
`any-light particle search' ALPS-I \cite{Ehret:2009sq,Ehret:2010mh} at DESY.
Proposals to enhance the sensitivity of this experiment even further, involve
an enlarged magnetic length \cite{Arias:2010bh} and the installation of a second cavity on the regeneration side, 
i.e., utilizing so-called `resonant-regeneration' \cite{Hoogeveen:1992uk,Hoogeveen:1990vq,Jaeckel:2007ch,Sikivie:2007qm}.
Higher energy photons \cite{Battesti:2010dm,Rabadan:2005dm} are used in other contexts.
Noteworthy, the latter concept could recently be verified experimentally
even for regeneration in the sub-quantum regime \cite{Hartnett:2011zd}.

\begin{figure}
\includegraphics[width=0.33\textwidth]{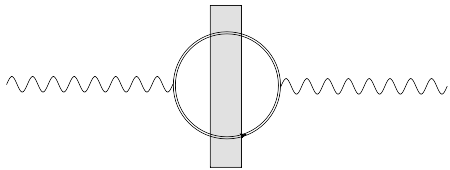} 
\caption{LSW scenario with virtual minicharged particles, also referred to as
  `tunneling of the 3rd kind', cf. \cite{Gies:2009wx}.  A photon can traverse
  a light-blocking barrier through a virtual fermionic or scalar
  particle-antiparticle loop. This process is possible at zero field, cf. \cite{Gies:2009wx}.
  Here we study the process in the presence of
  an external magnetic field. The dressed minicharged particle propagator, involving an
  arbitrary number of external field insertions, is represented by the solid
  double line.
}
\label{fig:t3k_full}
\end{figure}

In this paper we aim at investigating a different LSW scenario, namely the
`tunneling' of a barrier through in general \textit{virtual} particle-antiparticle 
intermediate states. Our main focus is on the process in the presence of an external magnetic field, cf. Fig.~\ref{fig:t3k_full}.
A compact version of our results has been given in \cite{Dobrich:2012sw}.

As the tunneling process with the virtual particle-antiparticle pair
complements both quantum mechanical tunneling and the tunneling of real particles as depicted in Fig.~\ref{fig:lsw_WISP},
it has been baptized `tunneling of the 3rd kind' \cite{Gies:2009wx}.
Although such a process is in principle also
possible with neutrinos, this standard model background is
highly suppressed due to the Fermi scale.

Although the photon-to-photon transition amplitude grows logarithmically
with decreasing mass of the minicharged particles
in the absence of an external field, zero-field
 `tunneling of the 3rd kind' cannot improve existing bounds of the 
minicharged particles' parameter space \cite{Gies:2009wx}.
By contrast, we show in this work that
current laboratory bounds on
minicharged fermions in the sub-meV regime can be significantly enhanced by an external magnetic field.
Most importantly, they can even outmatch the best current
model-independent cosmological bounds below minicharged fermion masses of $\mathcal{O} (10^{-4}) \mathrm{eV}$.

Our paper is organized as follows:
In Sect. \ref{sec:setting} the general formalism describing the LSW scenario depicted in Fig. \ref{fig:t3k_full} is outlined and the
photon-to-photon transition probability is derived.
Sect.~\ref{sec:poltensor} focuses on the explicit determination of the transition probability. This in particular requires an adequate analytical treatment of the photon polarization tensor.
Appropriate limiting cases which ultimately allow for a numerical evaluation of the transition probability are worked out.
In addition, analytic asymptotics in the most relevant limits are derived.
Specializing to a specific, experimentally feasible setup, in
Sect.~\ref{sec:bounds_chap} the corresponding exclusion plot in the fractional-charge mass plane are worked out and discussed in detail. 
The paper ends with conclusions and an outlook in Sect.~\ref{sec:conclusion}.

\section{Setting \label{sec:setting}}

Let us briefly review the basic setting as introduced in~\cite{Gies:2009wx} and extend it to account for an external magnetic field
$\vec{B}\neq0$. We adopt the metric convention $g_{\mu \nu}=\mathrm{diag}(-1,+1,+1,+1)$ of
 \cite{Gies:2009wx}, such that the momentum four-vector squared
reads $k^2= \vec{k}^2-\omega^2$.

We start with the effective Lagrangian describing photon propagation in a constant external magnetic field of strength $B=|\vec{B}|$ given by,
\begin{multline}
\mathcal{L}[A]= -\frac{1}{4} F_{\mu\nu}(x) F^{\mu\nu}(x) \\
- \frac{1}{2}\int_{x'}  A_\mu(x) \Pi^{\mu\nu}(x,x'|B) A_\nu(x')\ ,\label{eq:calL}
\end{multline}
where $\Pi^{\mu\nu}(x,x'|B)$ denotes the photon polarization tensor in the presence of the
external field, $F_{\mu \nu}$ the field strength tensor of the propagating
photon $A_\mu$, and $x$ a spatio-temporal four-vector.

As a result of their negligible coupling to matter, the wall does not affect the propagation of minicharged particles.
Hence, translational invariance is maintained for the minicharged particles and thus for the
polarization tensor at one-loop order. As long as the magnetic field is homogeneous
in the relevant spacetime region, this implies
$\Pi^{\mu\nu}(x,x'|B)=\Pi^{\mu\nu}(x-x'|B)$. 

Upon a variation of \Eqref{eq:calL} and a transformation to momentum space,
the following equation of motion is obtained
\begin{equation}
 \left(k^2 g^{\mu \nu} - k^\mu k^\nu + \Pi^{\mu\nu}(k|B) \right) A_\nu (k) =0 \label{eq:EOM_Poltensor} \ .
\end{equation}
The photon-to-photon transition probability depends on the polarization mode of the photons. Contrary to the zero-field situation,
there are three independent polarization modes in the presence of an external 
field, which we label by an index $p=1,2,3$.
As the vacuum speed of light in external
fields deviates from its zero-field value \cite{Heisenberg:1936qt,weisskopf,Schwinger:1951nm}, and the vacuum exhibits medium-like
properties, the occurrence of three (instead of two in the absence of an external field) independent polarization modes is not surprising. For instance, the corresponding polarization
effects induced by minicharged particles have been studied in
\cite{Gies:2006ca,Ahlers:2006iz}. Further diffractive effects in inhomogeneous
fields as also considered in quantum electrodynamics (QED) \cite{King} are conceivable and certainly worthwhile
to be studied.

Let us now briefly introduce the explicit structure of the photon polarization tensor in the presence of an external field.
Here we limit ourselves to the special case of a purely magnetic field \cite{Tsai:1974ap}. 
This naturally suggests a decomposition of the photon four-momentum $k^{\mu}$ in components parallel and 
perpendicular to the magnetic field vector $\vec{B}$. The angle $\varangle(\mathbf{B},\mathbf{k})$ is denoted by $\theta$. Without loss of generality $\vec{B}$ is assumed to point in $\vec{e}_1$ direction, and the following decomposition is adopted,
\begin{align}
 k^{\mu}&=k_{\parallel}^{\mu}+k_{\perp}^{\mu}\,,\nonumber\\
k_{\parallel}^{\mu}&=(\omega,k^1,0,0)\,,\quad k_{\perp}^{\mu}=(0,0,k^2,k^3)\,.
\end{align}
In the same manner, tensors can be decomposed, e.g., $g^{\mu\nu}=g_{\parallel}^{\mu\nu}+g_{\perp}^{\mu\nu}$. It is then convenient 
to introduce the following projection operators onto photon polarization modes,
\begin{align}
 P^{\mu\nu}_{1}=g_{\parallel}^{\mu\nu}-\frac{k_{\parallel}^{\mu}k_{\parallel}^{\nu}}{k_{\parallel}^2}\,, \quad{\rm and}\quad
 P^{\mu\nu}_{2}=g_{\perp}^{\mu\nu}-\frac{k_{\perp}^{\mu}k_{\perp}^{\nu}}{k_{\perp}^2}\,. \label{Pparperp}
\end{align}
Defining a third projector as follows,
\begin{align}
 P^{\mu\nu}_{3}=g^{\mu\nu}-\frac{k^{\mu}k^{\nu}}{k^2}-P^{\mu\nu}_{1}-P^{\mu\nu}_{2}\,, \label{P0}
\end{align}
the three projectors $P^{\mu\nu}_{p}$ obviously span the transverse subspace. 
Whereas two of these polarization modes can be continuously related to the photon polarization modes in the absence of an external magnetic field,
the third mode manifests itself in the presence of the external field only. 

Note that $P^{\mu\nu}_{1}$ and $P^{\mu\nu}_{2}$ have an intuitive interpretation, 
provided that $\vec{k}\nparallel\vec{B}$. Namely, they project onto photon modes 
polarized parallel and perpendicular to the plane spanned by the two vectors, $\vec{k}$ and $\vec{B}$. For $\vec{k}\nparallel\vec{B}$ these are the
polarization modes, that can be continuously related to those in the zero-field limit.
Remarkably, for the special alignment of $\vec{k}\parallel\vec{B}$, the situation is different. Here, the modes $2$ and $3$ can be continuously related to the 
two zero-field polarization modes.

With the help of Eqs.~(\ref{Pparperp}) and (\ref{P0}), the photon polarization tensor in a purely magnetic field can be decomposed as follows \cite{Dittrich:2000zu,Karbstein:2011ja},
\begin{equation}
 \Pi^{\mu\nu}(k) = \ \Pi_{1}(k)\,P^{\mu\nu}_{1}+\Pi_{2}(k)\,P^{\mu\nu}_{2}+\Pi_{3}(k)\,P^{\mu\nu}_{3}\,,
\label{eq:Pi_dec}
\end{equation} 
where the scalar functions $\Pi_{p}$ are the components of the polarization tensor in the respective subspaces.
Thus, dropping Lorentz indices, the equation of motion for photons in mode $p$,
$A_{p,\mu}= P_{p,\mu\nu}A^{\nu}$, resulting from \Eqref{eq:EOM_Poltensor} reads
\begin{equation}
\big(k^2 + \Pi_p(k|B) \big) A_{p}(k)=0\ .
\end{equation}

\begin{figure}
\includegraphics[width=0.35\textwidth]{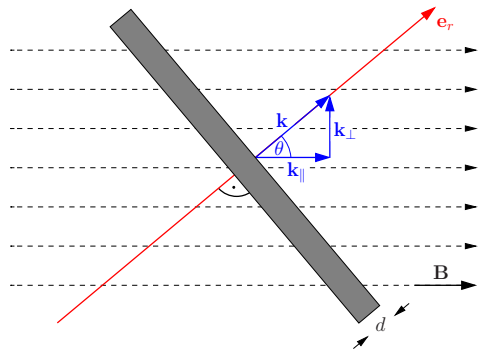} 
\caption{The probe photons are assumed to enter in ${\bf e}_r$ direction normal to the wall.
We look for regenerated photons behind the wall propagating along 
the line of propagation of the incident photons.
The angle between $\bf k$ and $\bf B$ is denoted by $\theta$.}
\label{fig:schief}
\end{figure}

As depicted in Fig.~\ref{fig:schief}, we assume the probe photons to propagate in ${\bf e}_r$ direction, normal to the light blocking barrier, i.e., $\vec k\parallel{\bf e}_r$.
The barrier extends from $r=0$ to $r=d$ along ${\bf e}_r$, and 
has infinite extent orthogonal to ${\bf e}_r$.

Introducing partial Fourier transforms of the photon field and the polarization tensor,
\begin{align}
A_p(r,\omega)&=\int\frac{{\rm d}\rm{k}}{2\pi}\,e^{ir\rm{k}}\,A_p(k)\ , \\
\Pi_p(r-r',\omega|B)&=\int\frac{{\rm d}\rm{k}}{2\pi}\,e^{{i}(r-r')\rm{k}}\,\Pi_p(k|B)\ , \label{eq:Fourier}
\end{align}
where $\rm{k}=|\bf{k}|$, the equation of motion becomes
\begin{equation}
\label{eq:j}
(\omega^2 + \partial_r^2) \,A_{p}(r,\omega) =
\underbrace{\int\mathrm{d} r' \,\Pi_p(r-r',\omega|B)\,A_{p}(r',\omega)}_{=:j_p(r,\omega|B)}  \,,
\end{equation}
where we have defined the fluctuation-induced current $j$.
Following~\cite{Gies:2009wx}, we choose reflecting boundary conditions on the front side of the barrier, $r=0$, for the incident
photons. Note that the calculation could easily be generalized to other boundary conditions. 
Accordingly, the induced current within and beyond the barrier reads
\begin{equation}
j_p(r>0,\omega|B) = \int\limits_{-\infty}^0 \mathrm{d} r''\, \Pi_p(r-r'',\omega|B)\, a(\omega)\, \sin
(\omega r'')\,, \label{eq:jright}
\end{equation}
where $a(\omega)$ denotes the amplitude of the incident photons. Moreover, the photon source is assumed to be sufficiently far away, such that the lower integration limit in \Eqref{eq:jright} can be formally sent to $-\infty$ 
(cf. comment at the end of this section). 

To obtain the photon-to-photon transition probability, the outgoing photon wave on the rear side of the barrier
has to be determined. 
For detector positions asymptotically far beyond the barrier at $r\gg d$, we find 
\begin{equation}
A^{\mathrm{out}}_p(r\gg d, \omega|B)= i \int\limits_d^\infty \mathrm{d} r'\, \frac{e^{i\omega (r-r')}}{2\omega} \,
j_p(r',\omega|B)\,, \label{eq:Aind}
\end{equation}
where we formally sent the upper integration limit to $+\infty$, employed the free outgoing Green's function for the operator
$(\omega^2+\partial_r^2)$ and restricted ourselves to the right-moving,
i.e., transmitted components of the outgoing photon field. This corresponds to
absorbing boundary conditions on the rear side of the wall at $r=d$.

Thus, combining \Eqref{eq:jright} and \Eqref{eq:Aind}, the photon-to-photon transition
probability can finally be written as 
\begin{multline}
\label{eq:transition}
P_{p,\gamma\to\gamma} = \lim_{r\to\infty} \left|\frac{A^{\rm out}_p(r,\omega)}{a(\omega)}\right|^2 \\
=\left|\,\int\limits^{\infty}_{d}\!\mathrm{d}r'\,\frac{e^{-i\omega r'}}{2\omega}\int\limits^{0}_{-\infty}\!\mathrm{d}r''\,
 \Pi_p(r'-r'')\sin(\omega r'')\,\right|^2,
\end{multline}
where we have dropped any explicit reference to the magnetic field $B$ for clarity. 

As $\Pi_p(r'\!-\!r'')$ receives its main contributions from relative distances $|r'\!-\!r''|$
of the order of the Compton wavelength $\sim1/m$ of the minicharged particles, and falls off rapidly for $|r'\!-\!r''|\gg1/m$,
it is justified to formally extend the integration ranges to $\pm\infty$, respectively.
With respect to an actual experimental realization this implies that the magnetic field has to be sufficiently homogeneous only
within a sphere of diameter $\gtrsim1/m$ centered at the intersection of the optical axis with the wall.

In the subsequent section we explicitly evaluate \Eqref{eq:transition}.
Our focus is on minicharged Dirac fermions. Remarks
on the rather different physics of scalar minicharged particles are postponed to
App. \ref{sec:mini_bosons}.

\section{Polarization tensor and transition probabilities \label{sec:poltensor}}

\subsection{Photon polarization tensor in momentum space \label{sec:poltensor_mom}}

In the presence of an external electromagnetic field, the photon polarization tensor in momentum space is explicitly known at one-loop accuracy.
Utilizing the propertime representation \cite{Schwinger:1951nm}, it is most
conveniently stated in terms of a double parameter integral \cite{Dittrich:2000zu,BatShab}, which in general cannot be tackled analytically (see, e.g., \cite{Dittrich:2000zu} 
and references therein).
Here we are interested in the special case of a purely magnetic field \cite{Tsai:1974ap} only. 
Several well-established approximations to the polarization tensor in a purely magnetic field have been worked out \cite{Tsai:1974fa,Tsai:1975iz,Shabad:1975ik,Baier:2009it}.
However, most approximations considered so far put special attention to
on-the-light-cone dynamics. Besides, they are in general limited to physical problems
that can be treated directly in momentum space, as their derivation involves constraints to a certain momentum regime. 
In contrast, our approach involves a Fourier transform to position space, cf. \Eqref{eq:Fourier}, and hence manifestly
requires knowledge about the full momentum regime.

To achieve this, we turn to the explicit expression of the photon polarization tensor in a purely magnetic field at one-loop accuracy (as reproduced in our notation in \cite{Karbstein:2011ja}).
The photon polarization tensor of standard quantum electrodynamics is adapted to minicharged particles by 
substituting the electron-positron loop for a minicharged particle-antiparticle loop. Formally, this amounts to a change of the coupling
$e \to \epsilon e$, with $\epsilon$ referring to a dimensionless fractional coupling, 
and an identification of $m$ with the mass of the minicharged particles.
Noteworthy, higher loop corrections to the photon polarization tensor at one-loop accuracy adapted to minicharged particles
are therewith suppressed by at least an additional factor of $\epsilon^2$.

We perform formal expansions in $\epsilon eB\to0$ and $\epsilon eB\to\infty$.
The combined quantity $\epsilon eB$ appears as the natural parameter, since the coupling of the external field $B$ to the minicharged particle line is $\epsilon e$.
The corresponding (dimensionless) physical expansion parameters, and the ranges of applicability of these expansions will be discussed on the level of the photon-to-photon transition amplitude below.

A perturbative expansion of the photon polarization tensor in powers of $\epsilon eB$, while retaining its full momentum dependence, is straightforward. It results in
\begin{align}
 \left\{
 \begin{array}{c}
\Pi_1\\
 \Pi_2\\
 \Pi_3
 \end{array}
\right\}=\Pi^{(0)}+
\left\{
 \begin{array}{c}
\Pi^{(2)}_1\\ 
\Pi^{(2)}_2\\
 \Pi^{(2)}_3
 \end{array}
\right\}+{\cal O}\left((\epsilon eB)^4\right)\,,
\label{eq:PI_pert}
\end{align}
were the upper index ${(n)}$ refers to the order in the expansion, i.e., denotes contributions of order $(\epsilon eB)^{n}$. In consequence of Furry's theorem, the expansion is in terms of even powers of $\epsilon eB$ only.
$\Pi^{(0)}$ is the photon polarization tensor in the zero-field limit,
\begin{equation}
\label{eq:pol_vac}
 \Pi^{(0)}=(k^2)^2\frac{\epsilon^2\alpha}{4\pi}\int\limits_0^1{\rm d}\nu\left(\frac{\nu^2}{3}-1\right)\frac{\nu^2}{\Phi_0}\ ,
\end{equation}
with
\begin{equation}
\Phi_0=m^2-{i}\eta+\frac{1-\nu^2}{4}k^2\ .
\label{eq:phi0}
\end{equation}
Here $\eta>0$ denotes an infinitesimal parameter
and $\alpha=e^2/(4\pi)$ is the fine-structure constant.
The contribution at ${\cal O}\left((\epsilon eB)^2\right)$ reads
\begin{align}
&\left\{
 \begin{array}{c}
\Pi^{(2)}_1\\ 
 \Pi^{(2)}_2\\
 \Pi^{(2)}_3
 \end{array}
\right\}
=-\frac{\epsilon^2\alpha (\epsilon eB)^2}{12\pi}\int\limits_{0}^{1}{\rm d}\nu\,\frac{(1-\nu^2)^2}{\Phi_0^2} \nonumber\\
&\times\!\left[
\left\{\!\!
\begin{array}{c}
\frac{-2}{1-\nu^2}\\
1\\
1
 \end{array}
\!\!\right\}
k_{\parallel}^2
+
\left(\left\{\!\!
\begin{array}{c}
1\\
\frac{5-\nu^2}{2-2\nu^2}\\
1
\end{array}
\!\!\right\}-\frac{1-\nu^2}{\Phi_0}\frac{k^2}{4}\right)
k_{\perp}^2
\right]\!\!.
\label{eq:PI_pert2}
\end{align}

Carrying out the expansion $\epsilon eB\to\infty$, while keeping the full momentum dependence, turns out to be more involved.
To leading order in the parameter $\epsilon eB$ we obtain \cite{DoebrichKarbstein,Karbstein:2011ja,Shabad:2003xy}
\begin{equation}
 \left\{
 \begin{array}{c}
\Pi_1\\
\Pi_2\\
\Pi_3
 \end{array}
\right\}=\frac{\epsilon^2\alpha \epsilon eB}{2\pi}\,{e}^{-\frac{k_{\perp}^2}{2\epsilon eB}}
\int\limits_{0}^{1}\!{\rm d}\nu
\left\{\!
 \begin{array}{c}
\frac{1-\nu^2}{\left.\Phi_0\right|_{\parallel}}k_{\parallel}^2\\ 
0\\
0
 \end{array}
\!\right\},
\label{eq:PI_strong}
\end{equation}
where we have introduced [cf. \Eqref{eq:phi0}]
\begin{equation}
 \left.\Phi_0\right|_{\parallel}=m^2-{i}\eta+\frac{1-\nu^2}{4}{k}_{\parallel}^2\,.
 \label{eq:phi0_largeB}
\end{equation}
Subleading corrections to \Eqref{eq:PI_strong} are at most logarithmic in $\epsilon eB$.
Hence, the strong field limit is dominated by $\Pi_1(k)$.  $\Pi_2(k)$ and $\Pi_3(k)$ are suppressed and start contributing at subleading order. 
Note that a factorization with respect to the momentum dependence,  $k_{\parallel}^{\mu}$ and $k_{\perp}^{\mu}$, is encountered in \Eqref{eq:PI_strong}.

Equations.~(\ref{eq:PI_pert}) and (\ref{eq:PI_strong}) are the
representations of the polarization tensor which serve as the basis of our
further analysis. 

\subsection{Towards the polarization tensor in position space\label{sec:poltensor_pos}}

Let us now explicitly implement the one-dimensional Fourier transform, \Eqref{eq:Fourier}, of the polarization tensor from momentum to position space. 
We first focus on the regime where the perturbative expansion in $\epsilon eB$ given in \Eqref{eq:PI_pert} holds. The Fourier transform of \Eqref{eq:PI_pert} involves terms of the general form
\begin{multline}
\int\limits_{-\infty}^{\infty}\frac{{\rm d}{\rm k}}{2\pi}\,e^{{i}(r-r'){\rm k}}\left[\frac{1-\nu^2}{\Phi_0}\right]^lP({\rm k}) \\
=4^l
\int\limits_{-\infty}^{\infty}\frac{{\rm d}{\rm k}}{2\pi}\frac{e^{{i}(r-r'){\rm k}}\,P({\rm k})}{\left[{\rm k}^2-\omega^2
\left(1-\frac{4m^2}{\omega^2}\frac{1-{i}\eta}{1-\nu^2}\right)\right]^l}\,,
\label{eq:Fourierblock_1}
\end{multline}
where $P({\rm k})$ denotes a polynomial in $\rm k$, and $l\in\mathbb{N}$.
Equation (\ref{eq:transition}) involves further integrations over the
spatial coordinates in the argument of $\Pi(r-r')$ 
such that the difference $r-r'$ is always non-zero and strictly
positive.

The evaluation of the $\rm k$ integral with the Residue theorem requires special care, as the location of the poles in the complex $\rm k$ plane 
depends on the further integration parameter $\nu$.
Noteworthy, for given $\omega>2m$, the quantity $\omega^2-\frac{4m^2}{1-\nu^2}$ exhibits a sign-change as a function of $\nu$.
Defining \cite{Gies:2009wx}
\begin{equation}
 \lambda=\sqrt{1-\frac{4m^2}{\omega^2(1-\nu^2)}}\ ,
\label{eq:lambda}
\end{equation}
and
\begin{equation}
 \kappa=\sqrt{\frac{4m^2}{\omega^2(1-\nu^2)}-1}\ ,
\label{eq:kappa}
\end{equation}
the poles of the integrand in \Eqref{eq:Fourierblock_1} in the complex $\rm k$ plane are located at
\begin{align}
 {\rm k}&=\pm{i}\omega\kappa\pm(1+{i})\eta\quad&{\rm for}\quad\quad\omega^2\leq\frac{4m^2}{1-\nu^2}\ , \nonumber\\
 {\rm k}&=\pm\omega\lambda\pm(1+{i})\eta\quad&{\rm for}\quad\quad\omega^2>\frac{4m^2}{1-\nu^2}\ .
\label{eq:k_poles}
\end{align}
This suggests the following decomposition of the $\nu$ integral in \Eqref{eq:PI_pert},
\begin{align}
 \int\limits_0^1{\rm d}\nu\,h(\nu) \to &-\hspace*{-3mm}
\int\limits_0^{\Re\sqrt{1-\frac{4m^2}{\omega^2}}}{\rm d}\lambda\left[\frac{{\rm d}\nu}{{\rm d}\lambda}\,h(\nu)\right]_{\nu=\sqrt{1-\frac{4m^2}{\omega^2(1-\lambda^2)}}}\nonumber\\
 &+\hspace*{-6mm}\quad\int\limits^{\infty}_{\Re\sqrt{\frac{4m^2}{\omega^2}-1}}{\rm d}\kappa
\left[\frac{{\rm d}\nu}{{\rm d}\kappa}\,h(\nu)\right]_{\nu=\sqrt{1-\frac{4m^2}{\omega^2(1+\kappa^2)}}}\,,
\label{eq:nu_decomposition}
\end{align}
where $h(\nu)$ denotes the integrand of the $\nu$ integral. With these preparations, the Fourier transform of \Eqref{eq:PI_pert} is implemented straightforwardly.
Subsequently we rely on the decomposition in \Eqref{eq:nu_decomposition}.
It is possible to distinguish two cases:
For $\omega\leq 2m$, there is only the $\kappa$ integral. For $\omega>2m$,
both the $\kappa$  and $\lambda$ integrals contribute. 
By virtue of the pole structure in the momentum integration, cf. \Eqref{eq:k_poles},
the integrand of the $\kappa$ part is exponentially damped with increasing $\kappa$, whereas the $\lambda$ part is oscillating as a function of $\lambda$.
Henceforth we label contributions due to $\kappa$ by ``$\leq$'', and contributions due to $\lambda$ by ``$>$''.

In the strong field limit with large $\epsilon eB$, \Eqref{eq:PI_strong}, the Fourier transform involves terms of the form
\begin{multline}
\int\limits_{-\infty}^{\infty}\frac{{\rm d}{\rm k}}{2\pi}\,e^{{i}(r-r'){\rm k}}\,\frac{1-\nu^2}{\left.\Phi_0\right|_{\parallel}}\,k_{\parallel}^2\,e^{-\frac{k_{\perp}^2}{2\epsilon eB}} \\
=
4\int\limits_{-\infty}^{\infty}\frac{{\rm d}{\rm k}}{2\pi}\frac{e^{{i}(r-r'){\rm k}}\,({\rm k}^2-\frac{\omega^2}{\cos^2\theta})\,e^{-\frac{{\rm k}^2\sin^2\theta}{2\epsilon eB}}}{{\rm k}^2-\frac{\omega^2}{\cos^2\theta}\left(1-\frac{4m^2}{\omega^2}\frac{1-{i}\eta}{1-\nu^2}\right)}\,.
\label{eq:Fourierblock_2}
\end{multline}
Due to the factor quadratic in $\rm k$ in the exponential, the integrand does no longer vanish in the limit ${\rm k}\to i\infty$, and the evaluation of the $\rm k$ integral becomes more complicated.
It is most conveniently performed with the help of the convolution theorem,
\begin{equation}
 \int\limits_{-\infty}^{\infty}\frac{{\rm d}{\rm k}}{2\pi}\,{\rm e}^{{i}{\rm k}r}\,g_1({\rm k})g_2({\rm k})\equiv\int\limits_{-\infty}^{\infty}{\rm d}\tilde r\,\hat g_1(\tilde r)\hat g_2(r-\tilde r)\,,
\label{eq:convolutionthm}
\end{equation}
where
\begin{equation}
 \hat g_i(r)=\int\limits_{-\infty}^{\infty}\frac{{\rm d}{\rm k}}{2\pi}\,{\rm e}^{{i}{\rm k}r}\,g_i({\rm k})\quad\quad (i=1,2)\,.
\end{equation}
Identifying
\begin{align}
 g_1({\rm k})&=\frac{{\rm k}^2-\frac{\omega^2}{\cos^2\theta}}{{\rm k}^2-\frac{\omega^2}{\cos^2\theta}\left(1-\frac{4m^2}{\omega^2}\frac{1-{i}\eta}{1-\nu^2}\right)}\,, \label{eq:g_1}\\
 g_2({\rm k})&= e^{-\frac{{\rm k}^2\sin^2\theta}{2\epsilon eB}}\,, \label{eq:g_2}
\end{align}
the Fourier transform of $g_1({\rm k})$ resembles that in \Eqref{eq:Fourierblock_1}. It can be performed with the Residue theorem.
The decomposition of the $\nu$ integral, \Eqref{eq:nu_decomposition}, remains favorable, but the poles in the complex ${\rm k}$ plane are shifted to
\begin{align}
 {\rm k}&=\pm\frac{i\omega\kappa}{\cos\theta}\pm(1+{i})\eta\quad&{\rm for}\quad\quad\omega^2\leq\frac{4m^2}{1-\nu^2}\ , \nonumber\\
 {\rm k}&=\pm\frac{\omega\lambda}{\cos\theta}\pm(1+{i})\eta\quad&{\rm for}\quad\quad\omega^2>\frac{4m^2}{1-\nu^2}\ .
\label{eq:k_poles_largeB}
\end{align}
The function $g_2({\rm k})$ does not depend on $\nu$. Its Fourier transform reads
\begin{equation}
\hat g_2(r)=\sqrt{\frac{\epsilon eB}{2\pi\sin^2\theta}}\,{e}^{-r^2\frac{\epsilon eB}{2\sin^2\theta}}\,.
\label{eq:g_2(r)}
\end{equation}
Therewith, the corresponding convolution integral on the right-hand side of \Eqref{eq:convolutionthm} can be rewritten in terms of error functions.
Let us however note that \Eqref{eq:g_2(r)} approaches the Dirac delta function in the formal limit $\frac{\sin^2\theta}{\epsilon eB}\to0$, via the heat kernel
\begin{equation}
\lim_{\tau\to0}\frac{1}{\sqrt{2\pi\tau}}\,{e}^{-\frac{r^2}{2\tau}}=\delta(r)\,. \label{eq:Diracdelta}
\end{equation}
This implies, that in the strong field limit, we are interested in here, we may approximate $\hat g_2(r)\simeq\delta(r)$. Using this 
in \Eqref{eq:convolutionthm}, the Fourier transform to be implemented in \Eqref{eq:Fourierblock_2} reduces to that of \Eqref{eq:g_1}.

\subsection{Towards the photon transition amplitude \label{sec:towards_fs}}

Next, we perform the two position-space integrals in
\Eqref{eq:transition}.
This can be achieved with the identity
\begin{multline}
 \int\limits_d^{\infty}{\rm d}r'\int\limits_{-\infty}^{0}{\rm d}r''\, \left[{i}(r'-r'')\right]^l e^{{i}\frac{\omega\lambda+i\eta}{\cos\theta}(r'-r'')}\sin(\omega r'')e^{-{i}\omega r'}\\
=\left(\frac{\cos\theta}{\omega}\right)^{2+l}\!(\partial_{\lambda})^l\!
\left[\!\frac{i\cos\theta\ e^{{i}\omega\left(\frac{\lambda}{\cos\theta}-1\right)d}}{(\cos\theta-i\eta-\lambda)^2(\cos\theta+\lambda)}\!\right],
\label{eq:identity}
\end{multline}
where $l\in\mathbb{N}_0$. Note that we explicitly account for the $i\eta$ prescription only in such terms where it is essential for the further calculation. The corresponding 
equation for $\kappa$ follows by substituting $\lambda\to{i}\kappa$.
After this substitution, the poles induced in the complex $\kappa$ plane are purely imaginary.
For angles $0\leq\theta<\frac{\pi}{2}$, not too close to $\theta=\frac{\pi}{2}$, these poles are located reasonably far away 
from the interval of integration, which corresponds to the real $\kappa$ axis.
In case of the $\lambda$ contribution, the situation is different. Here \Eqref{eq:identity} gives rise to poles at $\lambda=-\cos\theta$ and $\lambda=\cos\theta-i\eta$.
If the contour of the $\lambda$ integration is chosen
along the real $\lambda$ axis, and  $\theta$ is restricted as above, the first pole always lies well outside the interval of integration.
However, the latter pole {\it touches} the integration contour if 
\begin{equation}
 \cos\theta\leq\Re\sqrt{1-\frac{4m^2}{\omega^2}}\quad\leftrightarrow\quad\omega\sin\theta\geq 2m\,.
\label{eq:polcond}
\end{equation}
Let us remark that, sticking to a perturbative expansion in powers of
$\epsilon eB$, and assuming non-vanishing $\frac{2m}{\omega}$,
\Eqref{eq:identity} is employed for $\theta=0$ only.  This is a
consequence of the fact that all the Residues in \Eqref{eq:PI_pert}
arise from inverse powers of \Eqref{eq:phi0}.  As
condition~(\ref{eq:polcond}) cannot be fulfilled for $\theta=0$ and
$\frac{2m}{\omega}\neq0$, it is of no relevance in the perturbative
regime.

It is convenient to introduce dimensionless auxiliary functions $f_\leq$ and
$f_>$ as in~\cite{Gies:2009wx}, and to decompose the photon-to-photon transition probability in \Eqref{eq:transition} accordingly,
\begin{equation}
P_{\gamma \rightarrow \gamma}= \frac{\epsilon^4\alpha^2}{36 \pi^2} \big| f_{\leq} +  f_>  \big|^2.
\label{eq:def_f}
\end{equation}
The modulus in \Eqref{eq:def_f} allows us to drop global phase factors in the
definition of the auxiliary functions in the remainder. 

Before turning to the situation of a non-vanishing external magnetic field, let
us recall the auxiliary functions at zero field.
As already derived in \cite{Gies:2009wx}, for transversal photons we find
\begin{eqnarray}
f_\leq^{(0)} &=& \int_{\Re\sqrt{\frac{4m^2}{\omega^2}-1}}^\infty
{\rm d}\kappa\, \frac{ e^{-\omega d \kappa} }{i+\kappa}  \nonumber \\
&\phantom{=}&
\times
\frac{\sqrt{1+\kappa^2 - \frac{4m^2}{\omega^2}}\left( 1+\kappa^2 + \frac{2m^2}{\omega^2}\right)}{(1+\kappa^2)^{3/2}}\ ,
\label{eq:kappa_vac} \\
f_>^{(0)}  &=& \int_0^{\Re\sqrt{1- \frac{4m^2}{\omega^2}}} {\rm d}\lambda\,
\frac{ e^{i \omega d\lambda}} {1-\lambda} \nonumber  \\ 
&\phantom{=}&
 \times
 \frac{\sqrt{1-\lambda^2 -\frac{4m^2}{\omega^2}}\left( 1-\lambda^2 + \frac{2m^2}{\omega^2}
  \right) }{(1-\lambda^2)^{3/2}}
\ . \label{eq:lambda_vac}
\end{eqnarray}
The corresponding transition probability at zero field reads
\begin{equation}
\label{eq:P_vac_final}
P^{(0)}_{\gamma\to\gamma}= \frac{\epsilon^4\alpha^2}{36 \pi^2} \Big| f^{(0)}_\leq + f^{(0)}_>\Big|^2 .
\end{equation}
In addition to the results of \cite{Gies:2009wx}, we can provide for analytic asymptotics of \Eqref{eq:P_vac_final} in the limit where $\omega \gg2 m$ and $d\ll\frac{1}{m}\left(\frac{2m}{\omega}\right)^{-1}$ (cf. App.~\ref{sec:omega*d}).
Starting from $\lambda=0$, the integrand in \Eqref{eq:lambda_vac}
rapidly increases with $\lambda$.
The factor $\sqrt{1-\lambda^2-\frac{4m^2}{\omega^2}}$ guarantees that it is ultimately bent back 
and vanishes exactly at the upper integration limit.
This gives rise to a peak close to the upper integration limit.
The smaller the ratio $\frac{2m}{\omega}$, the closer the upper integration limit approaches the pole of the integrand at $\lambda=1$, 
and the more pronounced is the peak.
This motivates to approximate the following part of the integrand as,
\begin{equation}
\frac{(1-\lambda^2 + \frac{2m^2}{\omega^2})e^{i \omega d\lambda}}{(1-\lambda)(1-\lambda^2)^{3/2}}\approx\frac{1-\lambda + \frac{m^2}{\omega^2}}{\sqrt{2}(1-\lambda)^{5/2}}e^{i \omega d},
\label{eq:fexpansion_0}
\end{equation}
and the square root by
\begin{multline}
 \sqrt{1-\lambda^2-\tfrac{4m^2}{\omega^2}}\approx\sqrt{2}\left(\sqrt{1-\tfrac{4m^2}{\omega^2}}-\lambda\right)^{1/2}.
\label{eq:approx_0}
\end{multline}
This approximation retains the important feature that the integrand vanishes
at the upper integration limit, a behavior that would inevitably be spoiled
by a naive series expansion of the integrand in \Eqref{eq:lambda_vac} about $\lambda=1$. The resulting expression can be
integrated explicitly. Combining Eqs.~(\ref{eq:fexpansion_0}) and (\ref{eq:approx_0}), we obtain
\begin{multline}
\int\limits_{0}^{\sqrt{1-\frac{4m^2}{\omega^2}}} \!\!\!\mathrm{d} \lambda\, \frac{\left(1-\lambda + \frac{m^2}{\omega^2}\right)\left(\sqrt{1-\tfrac{4m^2}
{\omega^2}}-\lambda\right)^{1/2}e^{i \omega d}}{(1-\lambda)^{5/2}}\\
=-2\ln\left(\frac{2m}{\omega}\right){\rm e}^{i\omega d}+{\cal O}\left(\left(\tfrac{2m}{\omega}\right)^0\right).
\end{multline}

The integrand of the $\kappa$ contribution, \Eqref{eq:kappa_vac}, does not feature any pole close to the interval of integration and is highly suppressed as compared to \Eqref{eq:lambda_vac}.
Moreover, it decays exponentially with $\omega d \kappa$.
Hence, in the limit $\frac{2m}{\omega}\ll1$ and for $d\ll\frac{1}{m}\left(\frac{2m}{\omega}\right)^{-1}$ (cf. App.~\ref{sec:omega*d}), we obtain
\begin{align}
 f_>^{(0)}  &= -2\ln\left(\frac{2m}{\omega}\right){\rm e}^{i\omega d}+{\cal O}\left(\left(\tfrac{2m}{\omega}\right)^0\right), \nonumber\\
 f_\leq^{(0)} &= {\cal O}\left(\left(\tfrac{2m}{\omega}\right)^0\right),
\label{eq:f_vac_smallm}
\end{align}
and \Eqref{eq:P_vac_final} is well approximated by
\begin{equation}
\label{eq:P_vac_final_smallm}
P^{({\rm zero})}_{\gamma\to\gamma}\simeq \frac{\epsilon^4\alpha^2}{9 \pi^2}\ln^2\left(\frac{2m}{\omega}\right)\,,
\end{equation}
which is in agreement with the numerically determined asymptotics presented in Eq.~(3.17) of \cite{Gies:2009wx}.

\subsection{Transition amplitude for weak magnetic fields \label{sec:weak}}

\begin{figure}
\includegraphics[width=0.3\textwidth]{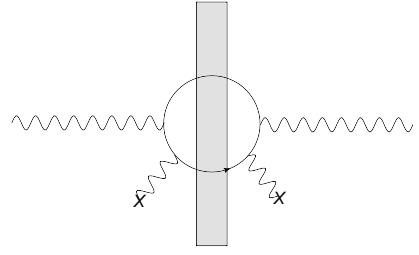} 
\caption{LSW scenario in the perturbative regime with two couplings to the external field.
This process merely serves as an illustration: as shown in Sect.~\ref{sec:weak}, the perturbative approximation 
ceases to hold far before the perturbative correction 
adds significant contributions to the zero-field result.}
\label{fig:t3k_pert}
\end{figure}

We now turn to the situation of a finite magnetic field.
It is instructive to determine the photon-to-photon transition probability in the perturbative regime first.
The leading perturbative correction to the zero-field limit is depicted schematically in Fig.~\ref{fig:t3k_pert}.

In this section, we limit ourselves to an angle of $\theta=0$. This significantly simplifies the expressions of the respective auxiliary functions, without neglecting any of their basic features. The results for non-zero angles follow straightforwardly from \Eqref{eq:PI_pert2}.

As elaborated in Sect.~\ref{sec:setting}, in this situation the modes labeled by $p=2,3$ correspond to the two photon propagation modes that can be continuously related to zero-field polarization modes.
Due to rotational invariance along the magnetic field vector, the components of the polarization tensor for these modes coincide for $\theta=0$ [cf. \Eqref{eq:PI_pert2}].
With the preparations in Sects.~\ref{sec:poltensor_pos} and \ref{sec:towards_fs}, it is straightforward to evaluate the corresponding transition amplitudes.
At order $(\epsilon eB)^2$, one finds
\begin{align}
\label{eq:kappa_small_B}
f_\leq^{(2)}&=-\left(\frac{\epsilon eB}{\omega^2}\right)^2\frac{4m^2}{\omega^2}\int_{\Re\sqrt{\frac{4m^2}{\omega^2}-1}}^\infty{\rm d}\kappa\,{e}^{-\omega d\kappa}\, \nonumber\\
&\quad\times\frac{i+2\kappa+d\omega\kappa(i+\kappa)}{\kappa^2(i+\kappa)^2(1+\kappa^2)^{3/2}\sqrt{1+\kappa^2-\frac{4m^2}{\omega^2}}}\, .
\end{align}
The corresponding function $f_>$ can be obtained from $f_\leq$ by substituting $\kappa\rightarrow -i\lambda$ and changing the integration boundaries 
accordingly [cf. \Eqref{eq:nu_decomposition}]. This yields
\begin{align}
\label{eq:lambda_small_B}
f_>^{(2)}&=\left(\frac{\epsilon eB}{\omega^2}\right)^2\frac{4m^2}{\omega^2}\int_0^{\Re\sqrt{1-\frac{4m^2}{\omega^2}}}{\rm d}\lambda\,{e}^{{i}\omega d\lambda} \nonumber\\
&\quad\times\frac{1-2\lambda-{i}d\omega\lambda(1-\lambda)}{\lambda^2(1-\lambda)^2(1-\lambda^2)^{3/2}\sqrt{1-\lambda^2-\frac{4m^2}{\omega^2}}}\,.
\end{align}
We briefly comment on the ranges of validity of these expressions.
A perturbative expansion in powers of $\epsilon eB$ is trustworthy when the dimensionful quantity $\epsilon eB$ is small in comparison to 
all the other quantities of the same dimension available.
For the setting considered here, this means that the following dimensionless ratios involving $\epsilon eB$ are small:
\begin{equation}
 \frac{\epsilon eB}{m^2}\ll1\quad{\rm and}\quad\frac{\epsilon eB}{\omega^2}\ll1\,. \label{eq:validity_pert}
\end{equation}
Note that we do not include a ratio involving the length scale $d$ in \Eqref{eq:validity_pert}. 
This is due to the fact that $d$ never appears alone, but always in the dimensionless combination $\omega d$. 

Let us have a closer look at Eqs.~(\ref{eq:kappa_small_B}) and (\ref{eq:lambda_small_B}).
In contrast to the zero-field limit, where \Eqref{eq:kappa_vac} is always finite, \Eqref{eq:kappa_small_B} is found to diverge
for $\omega=2m$ and all values of $d$.
As this divergence  or 'resonance phenomenon' is absent in the zero-field limit,
it can be considered as a genuine manifestation of the external field. 
Note that it has already been discussed by previous authors on
the level of the polarization tensor of QED, see,
e.g., \cite{shabad,Shabad:1972rg,Shabad:1975ik,Witte:1990}.
However, it becomes particularly relevant for minicharged
particles, since, in contrast to QED, the size of $m$ could
easily be smaller than the photon energies employed. 

The divergence in the photon-to-photon transition probability for $\omega=2m$ of course
signals a break down of unitarity in our calculation, as it would predict an arbitrarily large number of outgoing
photons for any small number of incoming photons. This unitarity violation is
a consequence of the idealized limit of a perfectly coherent infinite incoming
wave train. It would require a proper treatment, e.g., by taking the
finite line width of the laser into account. Yet, as the resonance may indeed lead to a strong
sensitivity to minicharged degrees of freedom fulfilling the condition $\omega=2m$, a careful analysis
seems highly worthwhile. However, in the present work we continue
to work in the perfectly coherent wave limit and ignore potential enhancements
arising from such resonances for our phenomenological conclusions, i.e., focus on the parameter space that can be firmly excluded even if the resonances would be smoothed out in an actual experimental realization.

A direct numerical evaluation of Eqs.~(\ref{eq:kappa_small_B}) and (\ref{eq:lambda_small_B}) becomes tedious for $\omega\geq2m$.
In this case the
integrand in \Eqref{eq:kappa_small_B} \textit{seems} to diverge at the lower integration
limit and that in \Eqref{eq:lambda_small_B} at both the lower and the upper
integration limits. However, the $i\eta$ prescription (cf. Sect.~\ref{sec:poltensor_pos}), 
assumed implicitly in Eqs.~(\ref{eq:kappa_small_B}) and (\ref{eq:lambda_small_B}), ensures that these
divergences do not lie on the integration contour.
Apart from the \textit{true} divergence at $\omega=2m$
discussed above, the encountered superficial divergences can be circumnavigated
with the help of integration by parts.

Using the following identity for $\omega\neq 2m$,
\begin{equation}
\label{eq:kpartint}
 \int \mathrm{d} \kappa \frac{1}{\kappa^2 \sqrt{1+\kappa ^2-\frac{4m^2}{\omega^2}}} = 
-\frac{\sqrt{1+\kappa
    ^2-\frac{4m^2}{\omega^2}}}{\left(1-\frac{4m^2}{\omega^2}\right) \kappa } +
C\ ,
\end{equation}
with $C$ denoting an integration constant, \Eqref{eq:kappa_small_B} can be rewritten as
\begin{multline}
\label{eq:smallB_olm_kappa}
f_<^{(2)}=  \left(\frac{\epsilon eB}{ \omega ^2}\right)^2\frac{\frac{4m^2}{\omega^2}}{1-\frac{4m^2}{\omega^2}}
\int_{\Re \sqrt{\frac{4m^2}{\omega^2}-1}}^{\infty} \mathrm{d} \kappa \,e^{-\omega d \kappa  }  \\
\times
\frac{\frac{i+8\kappa}{i+\kappa}+\omega d\left[\omega d  (1+\kappa^2)-2i+5\kappa \right]}{  (i+\kappa )(1+\kappa^2)^{5/2}} \sqrt{1+\kappa ^2-\tfrac{4m^2}{\omega^2}}
\,,
\end{multline}
and a surface term. Note that the lower label in \Eqref{eq:smallB_olm_kappa}
is ``$<$'', rather than ``$\leq$'', as the case $\omega=2m$ has been explicitly excluded.
Moreover, we have omitted the surface term of the integration by parts in
\Eqref{eq:smallB_olm_kappa}. It vanishes identically for $\omega<2m$. For
$\omega>2m$, it cancels with the corresponding surface term arising in an
analogous integration by parts of \Eqref{eq:lambda_small_B}. This can be easily
verified by taking into account the explicit $i\eta$ prescription (cf. Sect.~\ref{sec:poltensor_pos}).

Correspondingly, employing \Eqref{eq:kpartint} with $\kappa \rightarrow -i \lambda$,
and skipping the surface term, \Eqref{eq:lambda_small_B} becomes
\begin{multline}
\label{eq:smallB_olm_lambda}
f_>^{(2)} =  \left(\frac{\epsilon eB}{\omega ^2}\right)^2\frac{\frac{4m^2}{\omega^2}}{1-\frac{4 m^2}{\omega^2}}
 \int_0^{\Re\sqrt{1-\tfrac{4 m^2}{\omega ^2}}}\mathrm{d} \lambda\, e^{i \omega d \lambda } \\
\times
\frac{\frac{1-8\lambda}{1-\lambda}+\omega d\left[\omega d  (1-\lambda^2)  -i(2+5 \lambda )\right]}
{ (1-\lambda) (1-\lambda^2 )^{5/2}}\sqrt{1-\lambda^2-\tfrac{4 m^2}{\omega ^2}} \,.
\end{multline}
Equations~(\ref{eq:smallB_olm_kappa}) and (\ref{eq:smallB_olm_lambda}) are free of divergences in the intervals of integration,
and are thus perfectly suited for a numerical evaluation.

Finally, we combine the zero-field contributions, Eqs. (\ref{eq:kappa_vac}) and
(\ref{eq:lambda_vac}), with the leading perturbative corrections, Eqs.~(\ref{eq:smallB_olm_kappa}) and (\ref{eq:smallB_olm_lambda}), to
obtain the photon-to-photon transition probability, \Eqref{eq:def_f}, in the presence of weak magnetic fields,
\begin{equation}
\label{eq:P_small_B_final}
P^{({\rm weak})}_{\gamma\to\gamma}= \frac{\alpha^2 \epsilon^4}{36 \pi^2} 
\Big| f^{(0)}_\leq + f^{(0)}_> + f^{(2)}_< + f^{(2)}_> \Big|^2\,.
\end{equation}

Following the same reasoning as in the absence of an external field, we can provide for analytic asymptotics of \Eqref{eq:P_small_B_final} in the limit where 
$\omega \gg2 m$ and $d\ll\frac{1}{m}\left(\frac{2m}{\omega}\right)^{-1}$ (cf. App.~\ref{sec:omega*d}).
Now our focus is on \Eqref{eq:smallB_olm_lambda}. We approximate the following part of the integrand as,
\begin{multline}
\frac{\frac{1-8\lambda}{1-\lambda}+\omega d\left[\omega d  (1-\lambda^2)  -i(2+5 \lambda )\right]}
{ (1-\lambda) (1-\lambda^2 )^{5/2}}e^{ {i} \omega d \lambda } \\
= -\frac{7}{4\sqrt{2}} \frac{e^{{i}\omega d}}{(1-\lambda)^{9/2}} +{\cal O}\left((1-\lambda)^{-5/2}\right),
\label{eq:fexpansion}
\end{multline}
and use \Eqref{eq:approx_0} for the corresponding square root expression in \Eqref{eq:smallB_olm_lambda}.
To leading-order in $\frac{2m}{\omega}\ll1$, we obtain
\begin{multline}
-\frac{7}{4} \int\limits_{0}^{\sqrt{1-\frac{4m^2}{\omega^2}}}\!\!\! \mathrm{d} 
\lambda\, \frac{\left(\sqrt{1-\frac{4m^2}{\omega^2}}-\lambda\right)^{1/2}e^{{i}\omega d}}{(1-\lambda)^{9/2}}\\
= -\frac{32}{15} \left(\frac{2m}{\omega}\right)^{-6}e^{{i}\omega d}+{\cal O}\left(\left(\tfrac{2m}{\omega}\right)^{-4}\right).
\end{multline}
Hence, in the limit $\frac{\epsilon e B}{m^2}\ll1$, $\frac{2m}{\omega}\ll1$ and $d\ll\frac{1}{m}\left(\frac{2m}{\omega}\right)^{-1}$, \Eqref{eq:smallB_olm_lambda} is well approximated by
\begin{equation}
\label{eq:smallB_olm_lambda_approx}
f_>^{(2)} =  -\frac{2}{15}\left(\frac{\epsilon eB}{m^2}\right)^2e^{{i}\omega d}\left[1+{\cal O}\left(\left(\tfrac{2m}{\omega}\right)^2\right)\right].
\end{equation}
As in the zero-field case, \Eqref{eq:smallB_olm_kappa} is suppressed in comparison to \Eqref{eq:smallB_olm_lambda_approx} in the considered limit.
It is already clear at this point, that in the regime of validity of the weak field approximation, the transition probability for 
$\omega\gg 2m$ is still dominated by \Eqref{eq:f_vac_smallm}, which grows logarithmically with $\frac{2m}{\omega}\to0$.
Therefore, the leading contribution to \Eqref{eq:P_small_B_final} agrees with the leading contribution in the absence of an external field, and is given by \Eqref{eq:P_vac_final_smallm}.

\subsection{Transition amplitude for strong magnetic fields \label{sec:strong}}

Let us now aim at insights into the opposite regime, i.e., the limit of strong magnetic fields.
It is immediately obvious from \Eqref{eq:PI_strong}, that here the leading contribution arises from the $p=1$ component of the photon polarization tensor, whereas the components
$p=2,3$ contribute at subleading order only. 

Due to the overall exponential damping factor $\sim \exp\left
[-k_\perp^2/(2eB)\right]$ in \Eqref{eq:PI_strong} (cf. also~\cite{Melrose:1976dr,Dittrich:2000zu,Shabad:1975ik}), the maximum photon-to-photon transition probability attainable, is to be expected for small angles $\theta$.
Recall however, that in the strict limit $\theta=0$ the mode $p=1$ can no longer be continuously related to any of the photon polarization modes in the vacuum (cf. Sect.~\ref{sec:setting}).
It is not clear at all, if and how this mode can be excited for $\theta=0$, cf., e.g., the controversial viewpoints in \cite{Tsai:1975tw,Cover:1974ij}. 
Hence, our main focus is on a preferably small, but {\it finite} angle $\theta \gtrapprox 0$.
As discussed in detail in Sect.~\ref{sec:setting}, here the $p=1$ mode corresponds to transversal photons with polarization vector in the $(\vec{k},\vec{B})$ plane. It resembles an Alfv\'{e}n-like transversal mode.
The strict limit $\theta=0$ can be seen as constituting the asymptotics of an ideal experiment, allowing for an arbitrarily small but nonvanishing angle.
In contrast, the second (magneto-acoustic) polarization mode that can be continuously related to a photon polarization mode at zero field receives no dominant magnetic enhancement.

Recall, that in order to obtain \Eqref{eq:PI_strong}, we have considered the formal limit $\epsilon eB\to\infty$.
We now specify the physical parameter regime, where the corresponding expression for the photon-to-photon transition probability is trustworthy.
This turns out to be the case, if the following dimensionless ratios involving $\epsilon eB$ are large \cite{DoebrichKarbstein}:
\begin{equation}
 \frac{\epsilon eB}{m^2}\gg1\quad{\rm and}\quad\frac{\epsilon eB}{\omega^2\sin^2\theta}\gg1\,. \label{eq:validity_largeB}
\end{equation}
As in \Eqref{eq:validity_pert}, we do not include a ratio involving the length scale $d$ in \Eqref{eq:validity_largeB}. 
The length $d$ never occurs alone, but always in the dimensionless combination $\omega d$.
In contrast to \Eqref{eq:validity_pert}, the second ratio in \Eqref{eq:validity_largeB} involves a factor of $\sin^2\theta$ in the denominator. 
This is a direct consequence of the factorization with respect to the momentum dependence, encountered for the photon polarization tensor in the strong-field limit [cf. \Eqref{eq:PI_strong}].

Note that the conditions in \Eqref{eq:validity_largeB} also have an intuitive interpretation.
A magnetic field results in Landau level quantization of the momentum components perpendicular to the magnetic field vector $\bf B$.
On the level of the photon polarization tensor,
this  implies a factorization of the loop integral into a sum over discrete $\perp$ momentum components, and a continuous momentum integration for the $\parallel$ components.
The discrete $\perp$ momentum contributions can be reinterpreted in terms of an effective, level dependent mass.
Contributions of two successive Landau levels can be seen as differing in their effective masses squared by $\Delta m^2\sim \epsilon e B$.

Hence, the first condition in \Eqref{eq:validity_largeB} ensures that the effective mass of particles in higher Landau levels is significantly larger as compared to the zeroth one, where the mass is unscreened. 
The latter condition in \Eqref{eq:validity_largeB} ensures that the probability of real pair creation at higher Landau levels is small.

As already discussed in Sec.~\ref{sec:poltensor_pos}, for $\frac{\sin^2\theta}{\epsilon eB}\to0$ the Fourier transform of \Eqref{eq:g_2} becomes a representation of the Dirac delta function.
This limit is fully compatible with the parameter regime specified in \Eqref{eq:validity_largeB}.
For $\frac{\epsilon eB}{\sin^2\theta}\gg1$ an approximation of \Eqref{eq:g_2(r)} by the Dirac delta function is well justified, and further simplifies the calculation. 

Following the steps outlined in Sects.~\ref{sec:poltensor_pos} and \ref{sec:towards_fs}, for the $p=1$ mode in the strong-field limit, we obtain
\begin{multline}
\label{eq:lambda_large_Ba}
f_\leq^{(\mathrm{strong})}=-3i\cos^2\theta\,\frac{\epsilon eB}{\omega^2}\frac{4m^2}{\omega^2}\int_{\Re\sqrt{\frac{4m^2}{\omega^2}-1}}^{\infty}{\rm d}\kappa\ {e}^{-\frac{\omega d}{\cos\theta}\kappa}\\
\times\frac{\left(1+\kappa^2-\frac{4m^2}{\omega^2}\right)^{-1/2}}{\sqrt{1+\kappa^2}\,(\cos\theta-i\kappa)^2(\cos\theta+i\kappa)}\,,
\end{multline}
and
\begin{multline}
\label{eq:lambda_large_B}
f_>^{(\mathrm{strong})}=3\cos^2\theta\,\frac{\epsilon eB}{\omega^2}\frac{4m^2}{\omega^2}\int_0^{\Re\sqrt{1-\frac{4m^2}{\omega^2}}}{\rm d}\lambda\ {e}^{{i}\frac{\omega d}{\cos\theta}\lambda}\\
\times\frac{\left(1-\lambda^2-\frac{4m^2}{\omega^2}\right)^{-1/2}}{\sqrt{1-\lambda^2}\,(\cos\theta-{i}\eta-\lambda)^2(\cos\theta+\lambda)}\,.
\end{multline}
Let us emphasize that the structure of Eqs.~(\ref{eq:lambda_large_Ba}) and (\ref{eq:lambda_large_B}) is different from the analogous expressions in the weak-field regime.
Most strikingly, the factorization with respect to the momentum dependence in \Eqref{eq:PI_strong} induces $\cos\theta$ terms in the denominators [cf. also \Eqref{eq:k_poles_largeB}].
As discussed in Sec.~\ref{sec:towards_fs}, a pole in the interval of integration is induced in \Eqref{eq:lambda_large_B}, if the condition~(\ref{eq:polcond}) is met.

It is in general convenient to rewrite \Eqref{eq:lambda_large_B} by employing the same integration by parts as in the weak field regime, cf. \Eqref{eq:kpartint}. This yields
\begin{multline}
\label{eq:kappa_large_B}
f_<^{(\mathrm{strong})}=\frac{\epsilon eB}{\omega^2}\frac{4m^2}{\omega^2}\,\frac{3\cos^2\theta}{1-\frac{4m^2}{\omega^2}}\int_{\Re\sqrt{\frac{4m^2}{\omega^2}-1}}^{\infty}
{\rm d}\kappa\ {e}^{-\frac{\omega d}{\cos\theta}\kappa}\\
\times\frac{\frac{i\omega d \kappa}{\cos\theta}+\frac{i\kappa^2}{1+\kappa^2}-\frac{\kappa}{\cos\theta+i\kappa}-\frac{2i\cos\theta}{\cos\theta-i\kappa}}{\sqrt{1+\kappa^2}\,(\cos\theta-
i\kappa)^2(\cos\theta+i\kappa)}\sqrt{1+\kappa^2-\tfrac{4m^2}{\omega^2}}\,.
\end{multline}
For the $\lambda$ contribution, this particular partial integration is helpful only if the integrand does not feature any poles in the interval of integration, i.e., if 
$\cos\theta>\Re\sqrt{1-\frac{4m^2}{\omega^2}}$.
It results in
\begin{multline}
f_>^{(\mathrm{strong})}=\frac{\epsilon eB}{\omega^2}\frac{4m^2}{\omega^2}\,\frac{3\cos^2\theta}{1-\frac{4m^2}{\omega^2}}\int_0^{\Re\sqrt{1-\frac{4m^2}{\omega^2}}}{\rm d}
\lambda\ {e}^{{i}\frac{\omega d}{\cos\theta}\lambda}\\
\times\frac{\frac{i\omega d \lambda}{\cos\theta}+\frac{\lambda^2}{1-\lambda^2}-\frac{\lambda}{\cos\theta+\lambda}+\frac{2\cos\theta}{\cos\theta-i\eta-\lambda}}
{\sqrt{1-\lambda^2}\,(\cos\theta-{i}\eta-\lambda)^2(\cos\theta+\lambda)}\sqrt{1-\lambda^2-\tfrac{4m^2}{\omega^2}}\,.
\label{eq:lambda_large_Bb}
\end{multline}
As in Sect.~\ref{sec:weak}, these integrations by parts are
only possible for $\omega\neq2m$, and the transition probability diverges for $\omega=2m$
\cite{shabad,Shabad:1972rg,Shabad:1975ik,Witte:1990}. 
The surface terms in both the $\kappa$ and the $\lambda$ contributions
vanish by themselves.

For the situation $\cos\theta\leq\Re\sqrt{1-\frac{4m^2}{\omega^2}}$, it is more convenient to rewrite \Eqref{eq:lambda_large_Ba} in such a way that no explicit reference to the infinitesimal 
parameter $\eta$ is necessary. As the respective expression is rather lengthy, we postpone it to App.~\ref{app:lambda}.

The photon-to-photon transition probability in the strong-field limit is given by
\begin{equation}
\label{eq:P_large_B_final}
P^{(\mathrm{strong})}_{\gamma\to\gamma}= \frac{\epsilon^4\alpha^2}{36 \pi^2} 
\Big|f^{(\mathrm{strong})}_< + f^{(\mathrm{strong})}_> \Big|^2\,.
\end{equation}
In accordance with the previous sections, we aim at analytical insights into the limit where $\omega\gg2m$ and $d\ll\frac{1}{m}\left(\frac{2m}{\omega}\right)^{-1}$.

We first consider the regime, where
\begin{equation}
  \cos\theta>\Re\sqrt{1-\frac{4m^2}{\omega^2}}\quad\leftrightarrow\quad \omega\sin\theta< 2m\,.
\label{eq:polnichtcond}
\end{equation}
If we want to obey \Eqref{eq:polnichtcond} together with $\omega\gg2m$, the angle $\theta$ has to be very small. 
This motivates an expansion of \Eqref{eq:lambda_large_Bb} in powers of $\theta$,
\begin{multline}
f_>^{(\mathrm{strong})}=\frac{3\,\frac{\epsilon eB}{\omega^2}\,\frac{4m^2}{\omega^2}}{1-\frac{4m^2}{\omega^2}}
\int_0^{\Re\sqrt{1-\frac{4m^2}{\omega^2}}}{\rm d}\lambda\ \sqrt{1-\lambda^2-\tfrac{4m^2}{\omega^2}}\\
\times\frac{\lambda\left(i\omega d-\frac{1}{1+\lambda}\right)+\frac{1}{1-\lambda}\left(2+\frac{\lambda^2}{1+\lambda}\right)}
{(1-\lambda)^{5/2}(1+\lambda)^{3/2}}{e}^{{i}\omega d\lambda}\ +\ {\cal O}\left(\theta^2\right)\,.
\label{eq:lambda_large_Bb1}
\end{multline}
As the integrand does not feature poles in the interval of integration, any explicit reference to $\eta$ has been omitted here.
We approximate the following part of the integrand in \Eqref{eq:lambda_large_Bb1} by
\begin{multline}
 \frac{\lambda\left(i\omega d-\frac{1}{1+\lambda}\right)+\frac{1}{1-\lambda}\left(2+\frac{\lambda^2}{1+\lambda}\right)}{(1-\lambda)^{5/2}(1+\lambda)^{3/2}}{e}^{{i}\omega d\lambda} \\
= -\frac{5}{4\sqrt{2}} \frac{e^{{i}\omega d}}{(1-\lambda)^{7/2}} +{\cal O}\left((1-\lambda)^{-5/2}\right),
\label{eq:fexpansion_strong}
\end{multline}
and the square root as in \Eqref{eq:approx_0}.
Combining Eqs.~(\ref{eq:approx_0}) and (\ref{eq:fexpansion_strong}), we obtain
\begin{multline}
- \frac{5}{4} \int_{0}^{\sqrt{1-\frac{4m^2}{\omega^2}}} \mathrm{d} \lambda \frac{\left(\sqrt{1-\frac{4m^2}{\omega^2}}-\lambda\right)^{1/2}e^{{i}\omega d}}{(1-\lambda)^{7/2}}\\
= -\frac{4}{3} \left(\frac{2m}{\omega}\right)^{-4}e^{{i}\omega d}+{\cal O}\left(\left(\tfrac{2m}{\omega}\right)^{-2}\right)\,.
\end{multline}
Hence, in the strong field limit $\left\{\frac{\epsilon e B}{m^2},\frac{\epsilon e B}{\omega^2\sin^2\theta}\right\}\gg1$, and for 
$\frac{2m}{\omega}\ll1$ and $d\ll\frac{1}{m}\left(\frac{2m}{\omega}\right)^{-1}$ (cf. App.~\ref{sec:omega*d}), \Eqref{eq:lambda_large_Bb} is well approximated by
\begin{equation}
\label{eq:largeB_lambda_approx}
f_>^{({\rm strong})} = -\frac{\epsilon eB}{m^2}\,e^{{i}\omega d}\left[1+{\cal O}\left(\tfrac{4m^2}{\omega^2}\right)\right]+{\cal O}\left(\theta^2\right).
\end{equation}
Exactly as in the situation of vanishing and weak fields, \Eqref{eq:kappa_large_B} (the $\kappa$ branch) 
is suppressed in comparison to \Eqref{eq:largeB_lambda_approx} (the $\lambda$ branch) in the considered limit.
Therefore, the asymptotics for the photon-to-photon transition probability in the regime~(\ref{eq:polnichtcond}) is independent of $\theta$, and reads
\begin{equation}
 P^{({\rm strong})}_{\gamma\to\gamma}\simeq  \frac{\epsilon ^4\alpha ^2}{36 \pi ^2}\left(\frac{\epsilon eB}
{m^2}\right)^2 \,.
\label{eq:strong_asymp}
\end{equation}
It is also possible to extract an analytic asymptotics for $\omega\gg2m$ and $d\ll\frac{1}{m}\left(\frac{2m}{\omega}\right)^{-1}$ in the regime complementary to 
\Eqref{eq:polnichtcond}, characterized by \Eqref{eq:polcond}. This necessitates a careful analysis of \Eqref{eq:superlangda},
applying the same strategies as employed when deriving the other asymptotics in the limit $\omega\gg2m$ above. For the same reasons as
 discussed before, the $\kappa$ contribution \Eqref{eq:kappa_large_B} is negligible. We find (cf. App.~\ref{app:lambda})
\begin{multline}
\label{eq:largeB_lambda_approx2}
f_>^{({\rm strong})} = -3\frac{\epsilon eB}{\omega^2}\frac{4m^2}{\omega^2}\frac{(1+\cos\theta)\cos^2\theta}{\sin^4\theta}\,e^{{i}\frac{\omega d}{\cos\theta}}\\
\times\left[\ln\left(\frac{2m}{\omega}\right)+{\cal O}\left(\left(\tfrac{2m}{\omega}\right)^0\right)\right],
\end{multline}
wherefrom we obtain
\begin{align}
 P^{({\rm strong})}_{\gamma\to\gamma}&\simeq \frac{\epsilon^4\alpha^2}{4\pi^2} \,\frac{(1+\cos\theta)^2\cos^4\theta}{\sin^8\theta} \nonumber\\
 &\hspace*{1.2cm}\times\left(\frac{\epsilon eB}{\omega^2}\frac{4m^2}{\omega^2}\right)^2\ln^2\left(\frac{2m}{\omega}\right),
\label{eq:strong_asymp2a}
\end{align}
and for small angles $\theta$,
\begin{equation}
 P^{({\rm strong})}_{\gamma\to\gamma}\simeq \frac{\epsilon^4\alpha^2}{\pi^2} 
\,\frac{1}{\theta^8}\left(\frac{\epsilon eB}{\omega^2}\frac{4m^2}{\omega^2}\right)^2\ln^2\left(\frac{2m}{\omega}\right).
\label{eq:strong_asymp2}
\end{equation}

Let us provide for an intuitive explanation of the condition $\omega\sin\theta=2m$, separating the two regimes, defined by Eqs.~(\ref{eq:polcond}) and (\ref{eq:polnichtcond}). 
It is the on-set condition for real pair creation. To see this, it is illustrative to perform a Lorentz transformation along the direction of the magnetic field vector, 
such that the parallel momentum component $\vec k_{\parallel}$ of the photons becomes zero,
i.e., $\vec k_{\parallel}'=0$. Vector components orthogonal to the direction of the Lorentz boost remain unaltered.
In this particular reference system (denoted by $'$), the particles are still subject to a homogeneous external magnetic field. This comes about as a Lorentz boost
in the direction of the magnetic field does not induce an electric field.
However, the light-cone condition for photons now reads
\begin{equation}
 k^2=\vec k_{\perp}^2-(k'^0)^2=0 \,,
\end{equation}
and the on-set condition for real pair creation becomes
\begin{equation}
 k'^0=|\vec k_{\perp}|=2m \quad\leftrightarrow\quad \omega\sin\theta=2m \,.
\label{eq:pair}
\end{equation}

Going gradually from the regime $\omega\sin\theta<2m$ beyond this limit, i.e., to $\omega\sin\theta>2m$, 
we expect the photon-to-photon transition probability to drop, as real particle-antiparticle pairs are unlikely to reconvert into photons.

Note that the photon-to-photon transition probability diverges at the real pair creation threshold $\omega\sin\theta=2m$.
An analogous unitarity violation has been encountered already for $\omega=2m$. As discussed in detail in Sec.~\ref{sec:weak},
it can be cured by taking into account the finite line width of the laser.
Let us remark, that the divergence at $\omega\sin\theta=2m$ is not visible in any perturbative expansion in powers of $\epsilon eB$. 
Perturbative calculations account for a finite number of external field insertions in the loop diagrams only, and the intermediate
 minicharged particle lines are described by zero field propagators. Hence, translational invariance is not broken in the propagators 
[cf. \Eqref{eq:phi0} vs. \Eqref{eq:phi0_largeB}], and the poles in the complex $\rm k$ plane are not altered as compared to the zero field limit 
[cf. \Eqref{eq:k_poles} vs. \Eqref{eq:k_poles_largeB}]. 

{We can also intuitively motivate the scaling $\sim\theta^{-8}$ of \Eqref{eq:strong_asymp2} for small angles.
The strong field regime, characterized by \Eqref{eq:validity_largeB}, is compatible with sending the frequency $\omega$ to infinity, i.e.,
\begin{equation}
 \frac{2m}{\omega}\to0\,,\quad\frac{\epsilon eB}{\omega^2}\to0\,,
\end{equation}
while keeping the combined quantity $\omega\sin\theta\approx\omega\theta$ finite, 
{such that still $\epsilon eB/(\omega\theta)^2 \gg 1$}.
In this limit, the asymptotics~\eqref{eq:strong_asymp2} should also be valid for any {reasonable thickness} $d$ of the barrier (cf. Appendix~\ref{sec:omega*d}).
The only remaining dimensionless ratios to govern the photon-to-photon transition probability are thus given by 
\begin{equation}
 \frac{\epsilon eB}{m^2}\,,\quad\frac{\epsilon eB}{\omega^2\theta^2}\quad\text{and}\quad\frac{4m^2}{\omega^2\theta^2} \,.
\end{equation}
Taking into account that the leading term of the photon polarization tensor in the strong field limit, \Eqref{eq:PI_strong}, is linear in $\epsilon eB$
and expecting the transition probability to diminish with diminishing mass,
we can straightforwardly argue that the leading (polynomial) contribution of \Eqref{eq:strong_asymp2} should at least scale as $\theta^{-8}$:
\begin{equation}
 P^{({\rm strong})}_{\gamma\to\gamma}\sim\left|\frac{\epsilon eB}{\omega^2\theta^2}\frac{4m^2}{\omega^2\theta^2}\right|^2=\frac{1}{\theta^8}\left(\frac{\epsilon eB}{\omega^2}\frac{4m^2}{\omega^2}\right)^2\,.
\end{equation}
}

Finally, note that we have derived expressions for the photon-to-photon transition probability in all the parameter regimes of interest.
In particular we provided analytic asymptotics in the limit $\omega\gg2m$ and  $d\ll\frac{1}{m}\left(\frac{2m}{\omega}\right)^{-1}$ for the zero, weak and strong 
magnetic field limits. In all these cases, the leading contribution to the photon
transition amplitude turned out to depend on the thickness of the barrier $d$
via a global phase factor only, rendering the photon-to-photon transition probability independent of $d$.
This behavior can also be understood intuitively. The typical size of the particle-antiparticle loops along the magnetic field lines is of the order of the virtual particles'
Compton wavelength, i.e., $\sim m^{-1}$. In the small mass limit, $\omega\gg2m$ and $d\ll\frac{1}{m}\left(\frac{2m}{\omega}\right)^{-1}$, and for small $\theta$,
we are in a situation where $d$ is significantly smaller than the Compton wavelength of the minicharged particles. 
Hence, the interval of width $d$ which is not available for photons to split or recombine into minicharged particles is negligible, and the width of the barrier 
does not affect the asymptotics of the transition probability.
As described above, beyond the threshold for real particle creation in a magnetic field, the asymptotics is governed by 
the conversion of photons into real minicharged particles. As these real particles are not reconverted into photons, 
they do not contribute to the photon-to-photon transition probability. Of course, this process also does not result in any $d$ dependence.

\section{A laboratory experiment for LSW with virtual minicharged particles \label{sec:bounds_chap}}

\subsection{Comparison of LSW scenarios with minicharged fermions}

\begin{figure}
\includegraphics[width=0.4\textwidth]{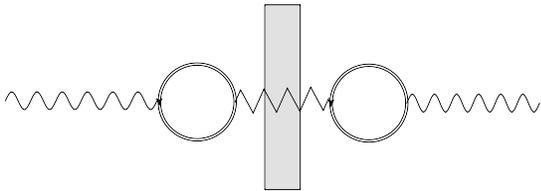}
                {\caption{LSW scenario with hidden photon and minicharged
                    particles. A photon (wiggly line) can oscillate into a
                    hidden photon (zig-zag-line) through a minicharged particle
                    loop. After traversing the wall, the hidden photon can be
                    reconverted into a photon by the reverse
                    process. From this scenario bounds on the
                    coupling $\epsilon$ can be deduced for fixed hidden photon to
                    minicharged particle coupling.  Note that here the external magnetic
                    field is not necessary, but favors the conversion process.}
\label{fig:MCP_hiddenph}}
\end{figure}

Being the currently most sensitive laboratory LSW experiment, the ALPS
collaboration already provides bounds \cite{Ehret:2010mh} on minicharged
particles through the LSW scenario depicted in Fig.~\ref{fig:MCP_hiddenph}: In
models where the fractional charge of the minicharged particles arises from the
gauge-kinetic mixing of the standard model $U(1)$ with an additional
hidden-sector $U(1)_{\mathrm{H}}$, `light shining through walls' is
possible if photons traverse a light-blocking barrier through a real hidden-photon
intermediate state \cite{Ahlers:2007rd,Ahlers:2007qf}. 

Obviously, the ALPS bounds on minicharged particles rely on the existence of
hidden photons. {\it Per se}, this is not unsatisfactory as many models with minicharged particles
employ a kinetic mixing mechanism \cite{Okun:1982xi,Holdom:1985ag}, such that
minicharged particles often come hand in hand with hidden photons. On the other hand,
the scenario as depicted in
Fig.~\ref{fig:MCP_hiddenph} (the external magnetic field is not
mandatory for the process, but amplifies it) does not give direct limits on
$\epsilon$ but rather only on a \textit{combined} parameter that involves
$\epsilon$ together with the minicharged particle to hidden-photon coupling, see,
e.g., \cite{Ahlers:2007rd}.  By contrast, the LSW scenario with virtual
minicharged particles, cf. Fig.~\ref{fig:t3k_full}, has the potential to provide
\textit{direct} limits on $\epsilon$.  For this reason it will be instructive
to compare the discovery potential of our setup to the
currently best laboratory exclusion limits on minicharged particles \cite{Ahlers:2007qf}, derived from PVLAS
\cite{Zavattini:2007ee} polarization measurements. These are slightly less
stringent in comparison to the ALPS LSW data, but do not rely on a hidden-photon
intermediate state.
Lastly note that, ALPS adopts $\theta= \varangle(\vec{B},\vec{k})=\pi/2$ as this both
maximizes the transition probability for standard processes involving
axion-like particles, and facilitates the discrimination of scalar and
pseudoscalar particle species, see, e.g., \cite{Ehret:2010mh}. By contrast, for
magnetically amplified `tunneling of the 3rd kind' a choice of
$\theta\approx0$ is favorable.

\subsection{Experimental parameter set and numerical implementation \label{sec:expparamset}}

In the following, we specify several details of a realistic setup with
rather conservatively chosen parameters in order to illustrate the
discovery potential of our LSW scenario. {To be specific, we
  mainly use parameters of the ALPS experiment as a realistic
  example. For a dedicated experiment, we expect that the optics
  design could easily be optimized.}

As in ALPS-I, {we consider} light of a frequency doubled standard
laser light source, $\omega=2.33 \ \mathrm{eV}$ ($\lambda=532
\mathrm{nm}$)\footnote{\label{fn:wavelength}Note that for the
  employment of a regeneration cavity, it can be favorable to use the
  laser in the fundamental mode, i.e., $\lambda=1064\mathrm{nm}$
  ($\omega=1.165$eV).  This possibility is also taken into account in
  Fig.~\ref{fig:excl_lhs} for $\theta=0.001^{\circ}$.}, {which} is fed into an
optical resonator cavity of length $L{=8.6{\rm m}}$ to increase
the light power available for minicharged particle production.
{The resonator uses a plano-concave design with one plane mirror
  and a curved one with radius of curvature $R=15{\rm m}$.  
The incident laser light is coupled into the resonator via the curved 
mirror and is directed towards the plane mirror, mounted right in front 
of the barrier. In this way, a stable resonator mode is built up in 
between the two mirrors \cite{Ehret:2009sq}.  For Gaussian beams, the beam waist
  $w_0$ in such a resonator is given by \cite{Siegman},
\begin{equation}
 w_0^2=\frac{\lambda}{\pi}\sqrt{L(R-L)}\,.
\end{equation}
For our parameters, $w_0\approx1.12{\rm mm}$.
The beam waist corresponds to the radius of the laser beam on the surface of the plane mirror.
Within the resonator, the radius $w$ of the laser beam increases as a function of the distance $D$ to the plane mirror \cite{Siegman},
  \begin{equation}
 w(D)=w_0\sqrt{1+\left(\frac{\lambda D}{\pi w_0^2}\right)^2}.
\end{equation}
Thus, whereas the photons are perfectly aligned with the optical axis on the surface of the plane mirror, i.e., at $D=0$, the maximum angular deviation from the optical axis $\Delta\theta$ in the resonator increases with $D$,
\begin{align}
 \Delta\theta(D)&=\arctan\left(\frac{\partial w(D)}{\partial D}\right) \nonumber\\
 &=\arctan\left[\left(\frac{\lambda}{\pi w_0}\right)^2\frac{D}{w(D)}\right]. \label{eq:DeltaTheta}
\end{align}
}
The smallest resolvable minicharged particle mass is
determined by several experimental constraints.

Firstly, it is limited by the extent of the homogeneous external magnetic field. More
specifically, the spatial and temporal homogeneity of the magnetic field
should be comparable to or larger than the Compton wavelength of the minicharged
particles.
In addition, our setup requires that the direction of the laser beam and accordingly its 
cavity can be adjusted such that they are almost perfectly aligned with 
the magnetic field lines. As emphasized above, this discriminates our experimental needs from those of existing LSW searches such as ALPS,
which employs a dipole magnet with $\theta= \varangle(\vec{B},\vec{k})=\pi/2$.  In consequence, 
for our purposes, accelerator dipole magnets, as used in ALPS, do not seem favorable, as there the existing magnet bore extends transversely to the magnetic field lines.
The ideal magnet in this tunneling experiment with virtual minicharged particles is 
presumably a solenoid or Helmholtz coil magnet\footnote{Notably, next generation heliosopes and direct 
dark matter searches suggest
the employment of large scale toroidal magnets, cf. \cite{Baker:2011na,Irastorza:2011gs}, which offer a large magnetic field region of the order of several meters.
For our LSW scenario, the employment of such a magnet would also be possible, particularly as the ATLAS B0 test coil 
allows for a cavity setup along the magnetic field lines. This would open up the possibility to probe minicharged particles with masses down to $\mathcal{O}(10^{-8})$eV.}.

As a suitable magnet we have identified a presently unused ZEUS compensation solenoid \cite{Dormicchi:1991ad} available at DESY.
It features a bore of $0.28$m diameter and $1.20$m length and provides a field strength of $B=5$T. The field points along the bore,
and is assumed to be adequately aligned on the solenoid's axis (accurate alignment studies of magnetic field lines relative to gravity have, e.g., been performed in \cite{Meinke:1990yk} for a HERA dipole magnet).
The field strength near the center of the solenoid is expected to be sufficiently homogeneous at least over a typical extent of the order of the bore's diameter.
The wall is installed in the center of the bore and the back end of the cavity extends into the bore. 
The angle $\theta$ is adjusted by tilting the entire optics assembly relative to the solenoid's axis.
As discussed below \Eqref{eq:transition}, the length
scale over which the field can be considered as approximately homogeneous
should be comparable to or larger than the Compton wavelength of the minicharged
particle. Thus, with the ZEUS compensation solenoid, access to minicharged particle masses
down to  $m \gtrsim 7\times 10^{-7}\mathrm{eV}$ {-- corresponding to a Compton wavelength of $0.28{\rm m}$ --} is granted.
As the minicharged particles interact extremely weakly with the
magnet material and all the other parts of the setup, the diameter of the beam tube and other parameters are not relevant. 

Secondly, as discussed above, the smallest resolvable minicharged particle mass is also limited by the pair production threshold $\omega \sin\theta=2m$.
Note, however, that beyond the pair production threshold, instruments sensitive to real minicharged particles could possibly be combined with our proposal
for an increased sensitivity also in this region, see, e.g. \cite{Gies:2006hv}. In the present study, we do not consider such options.

In practice, the relation for the pair production threshold directly constrains the smallest resolvable mass via the smallest experimentally realizable angle $\theta$.
On the one hand, the smallest achievable angle depends on the magnet bore, as the generation cavity (and possibly also the regeneration cavity, see below) 
needs to fit into the magnet bore and thus may not exceed a certain size. 
On the other hand, it must be ``large enough'' as to allow for a precise geometric alignment.
Below, we will give results for three exemplary angles, namely for $\theta=0.1^{\circ}, 0.01^{\circ}$ and $\theta=0.001^{\circ}$.
Given a precise alignment of the magnetic field lines {within a diameter of $\cal O({\rm mm})$ about} the solenoid's axis {-- recall, that $w_0\approx1.12{\rm mm}$ --}, the uncertainty in the
  adjustment of $\theta$ is expected to be dominated by the  {maximum angular deviation from the optical axis in the resonator} $\Delta\theta$. 
{For $D=0.6{\rm m}$, i.e., half the length of the ZEUS compensation solid, \Eqref{eq:DeltaTheta} yields $\Delta\theta\approx0.0007^\circ$. Given this small uncertainty in the alignment, angles down to $\theta=0.001^{\circ}$ should indeed be experimentally feasible. {In \cite{Dobrich:2012sw}},  we have {even} overestimated this uncertainty.}

In addition, we comment on the thickness $d$ of the light-blocking barrier, which is assumed to be perfectly reflecting, cf. Sect. \ref{sec:setting}.
For ALPS, the barrier is realized by a high-quality steel block of width
$d=1.8\mathrm{cm} \mathrel{\widehat{=}} 9.1 \times 10^{4} \mathrm{eV}^{-1}$ \cite{Knabbe}: As for ALPS the barrier is assumed to be traversed
by real particles, the photon-to-photon transition probability does not depend on the thickness of the wall in the limit of vanishing WISP-to-matter coupling.

By contrast, our tunneling scenario exhibits an intrinsic sensitivity on
the wall thickness: For minicharged particle masses corresponding to Compton wavelengths
smaller than $d$, the tunneling process is obstructed,
whereas for Compton wavelengths substantially larger than $d$, the photon-to-photon transition probability even becomes
independent of $d$, see also below.  To
demonstrate this, we provide exclusion limits for two different thicknesses of the
wall, namely $d=1 \mathrm{mm} \mathrel{\widehat{=}} 5.1 \times 10^{3}
\mathrm{eV}^{-1}$ and $d=1 \mathrm{\mu m} \mathrel{\widehat{=}} 5.1
\mathrm{eV}^{-1}$, which are most easily numerically accessible. 
However, let us stress that exploring large Compton wavelengths,
i.e., small minicharged particle masses with the considered scenario is experimentally also possible with thicker barriers,
as, e.g., used in the ALPS experiment.
{For minicharged particles of larger masses, the experimental challenge can be greater, as the minimization of the barrier thickness can become an important handle to maximize the transition probability.
As the discovery potential of our experiment is most promising in the small mass regime, no difficulties in the ``wall''-design should arise.}
Technically, the numerical evaluation of
the auxiliary function $f_>$ becomes considerably more involved for thicker barriers, see below.
For this reason, we limit the numerical evaluation to the $d$ values stated above. 

The asymptotics derived for the photon-to-photon transition probability, Eqs.~(\ref{eq:strong_asymp}) and (\ref{eq:strong_asymp2a}), are extremely helpful here.
They do not feature any explicit dependence on $d$, and are valid in the parameter regime where $\omega\gg 2m$, and for wall 
thicknesses $d\ll\frac{1}{m}\left(\frac{2m}{\omega}\right)^{-1}$. It is convenient to restate these findings also in a slightly different way:
For a given laser energy $\omega$ and wall thickness $d$, the asymptotics provide us with the correct result of the photon-to-photon transition 
probability for all minicharged particle masses $m\ll\left\{\frac{\omega}{2},\sqrt{\frac{\omega}{2d}}\right\}$.  
As demonstrated in Sec.~\ref{sec:bounds} below, this will immediately allow us to provide results for wall thicknesses of ${\cal O}({\rm cm})$, as, e.g., 
employed in ALPS, without the need for any additional numerical evaluations.

We proceed to the evaluation of the observable in the proposed LSW scenario,
which is the outgoing photon rate behind the light-blocking barrier:
\begin{equation}
\label{eq:conv}
n_{\mathrm{out}} = \mathcal{N} \  n_{\mathrm{in}} \  P_{\gamma\to\gamma} \ .
\end{equation}
The lever arm for probing small minicharged particle couplings is the large number of
incoming photons $ n_{\mathrm{in}}$ on the right hand side of \Eqref{eq:conv}
vs. -- in principle -- single-photon detection for $ n_{\mathrm{out}}$. 
In addition, we introduce a variable $\mathcal{N}$,
which accounts for the option of installing a regeneration cavity on the
 right hand side of the wall. 
Without a cavity on the regeneration side,
one has $\mathcal{N}=1$.

Inspired by the experimental possibilities at ALPS, we
assume an ingoing-to-outgoing photon ratio $ n_{\mathrm{in}}/ n_{\mathrm{out}} =10^{25}$, 
which already accounts for additional parameters of the experimental realization, i.e., the use of a front-side cavity, the
effective detector sensitivity and the running time, cf. \cite{Ehret:2010mh}.

Lastly, we comment on the numerical evaluation of \Eqref{eq:conv}.
Herein a difficulty is the numerical evaluation of the auxiliary function $f_>$, contributing to the photon-to-photon transition 
probability $P_{\gamma\to\gamma}$.

As a function of the integration variable $\lambda$, the integrand in $f_>$ generically exhibits a highly oscillatory behavior (depending on
the parameter set $d,\omega,m$) in all relevant limits (with or without external field): Rescaling the integration variable as $\lambda \rightarrow
\tilde{\lambda} \sqrt{1-\frac{4 m^2}{\omega^2}}$,  the numerical evaluation
becomes increasingly difficult in the vicinity of the upper integration boundary
$\tilde{\lambda} \lessapprox 1$ for larger values of the oscillation frequency
$d \sqrt{\omega^2 - 4 m^2}$: For thicker walls and masses fulfilling the relation
$\omega\gg 2m$, the frequency of the oscillations is always very large. It becomes
only slightly smaller for minicharged particle masses in the regime $\omega \gtrsim 2m$.
In addition also the amplitude of the oscillation becomes very large near
$\tilde{\lambda} \lessapprox 1$, especially for small $m$.  To
circumnavigate this problem, it is helpful to split the $\tilde{\lambda}$
integral at an appropriate cutoff value
$\tilde{\lambda}_\mathrm{c}(d,\omega,m)$ which depends on both the set of variables used
and on the employed numerical integration
routine.  Substituting $\tilde{\lambda} \rightarrow 1/R$ above the cutoff, we can
then rewrite the integral as: $\int_0^1 \mathrm{d} \tilde{\lambda} =
\int_0^{\tilde{\lambda}_\mathrm{c}} \mathrm{d} \tilde{\lambda} +
\int_1^{1/\tilde{\lambda}_\mathrm{c}} \mathrm{d} R/R^2$.
Thereafter, the numerical routine is capable of treating the problematic region.

In contrast to $f_>$, the integrand of $f_<$ is never oscillating, but always
decaying as a function of the integration variable $\kappa$, cf. Sect. \ref{sec:poltensor}. Nevertheless,
for the numerical evaluation, a similar rescaling procedure,
$\kappa \rightarrow \tilde{\kappa} \sqrt{\frac{4 m^2}{\omega^2}-1}$  for $\omega<2 m$ is consistent.
For $\omega>2m$, the numerical evaluation of
the $\kappa$ integral is straightforward.

Let us emphasize, that as we have worked out asymptotics for all relevant limits, we can easily obtain estimates for different parameter sets, without having to go through a full
numerical analysis.

\subsection{Prospective exclusion bounds for minicharged fermions \label{sec:bounds}}

\begin{figure}
\includegraphics[width=0.49\textwidth]{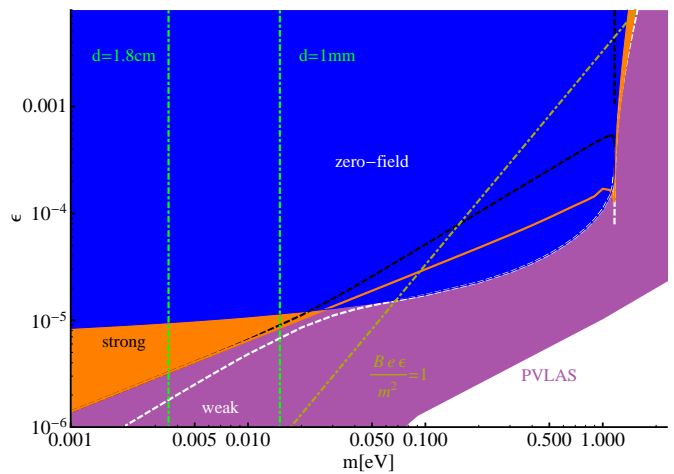} 
\caption{
  Exclusion bounds in the fractional-charge mass plane for `large' minicharged particle masses. Without an external field, the uppermost area, depicted in blue (dark gray), can be excluded.
  The white dashed line is determined with \Eqref{eq:P_small_B_final}. It delimits the excluded area from below.
  As the underlying weak field approximation is limited to $\frac{\epsilon e B }{m^2} \ll1$,
  the corresponding bounds can only hold to the right of the yellow dot-dashed line, representing $\frac{\epsilon e B }{m^2} =1$,
  where they do not significantly improve the zero-field bounds.
  For any finite magnetic field, however, a divergence is found in the photon-to-photon transition probability for $\omega=2m$ (cf. main text).
  The area above the orange line, particularly the orange (light gray) area in the lower left labeled by `strong', can be excluded with \Eqref{eq:P_large_B_final},
  valid for $\left\{\frac{\epsilon e B }{m^2},\frac{\epsilon e B }{\omega^2\sin^2\theta}\right\}\gg 1$, when
  assuming angles $\theta\leq0.1^{\circ}$ and using a wall thickness of $d=1\mathrm{\mu m}$ for easy numerical evaluation.
  The black dashed line defines the area excluded when setting $d=1\mathrm{mm}$ instead.
  It can be seen that the bounds become independent of $d$ below $m\approx 5\times 10^{-3} \mathrm{eV}$.
  Notably, the orange area is also perfectly reproduced when employing the analytic asymptotics~(\ref{eq:strong_asymp}) instead of \Eqref{eq:P_large_B_final}. 
  The green, dash-dotted vertical lines depict the condition $d=\frac{1}{m}\left(\frac{2m}{\omega}\right)^{-1}$ (cf. App.~\ref{sec:omega*d}) for $d=1$mm and $d=1.8$cm.
  For a given $d$, the asymptotics~(\ref{eq:strong_asymp}) become trustworthy to the left of the respective line.
  This theoretical prediction is perfectly compatible with the convergence of the black and the orange lines to the left of the $d=1$mm line.
  Hence, most importantly, all exclusion bounds to the left of the $d=1.8$cm (the wall thickness at ALPS) line are also valid for wall thicknesses of up to
 at least $d=1.8$cm,
  even though the respective numerical calculations employ $d=1\mu$m.
  As a reference, the parameter regime excluded by the PVLAS experiment is depicted in purple (gray).}
\label{fig:excl_rhs}
\end{figure}

\begin{figure}[h!]
\includegraphics[width=0.49\textwidth]{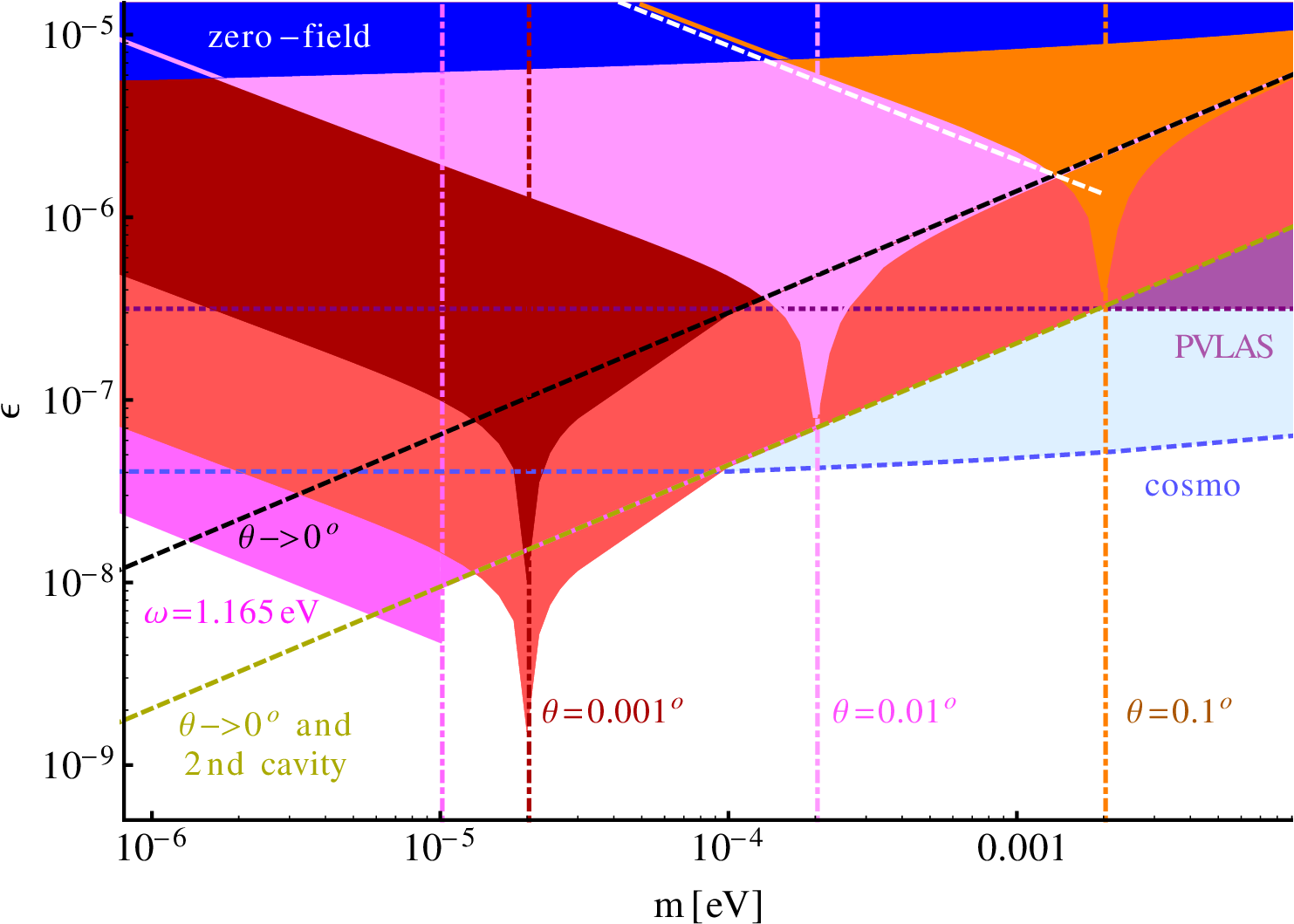} 
\caption{Exclusion bounds in the fractional-charge mass plane for `small' minicharged particle masses, valid for wall thicknesses up to at least $d=1.8$cm. 
  All results besides the zero-field bounds are derived numerically from \Eqref{eq:P_large_B_final}. The dark blue (darkest gray) area depicts the
  parameter regime that can be excluded in the zero-field limit, cf. also Fig.~\ref{fig:excl_rhs}. The orange (light gray), cone-shaped area on the very right
  depicts the parameter space that can be excluded when choosing an angle of $\theta=0.1^{\circ}$.
  Thereafter, from right to left, pink (gray) and dark red (dark gray) shaded cones correspond to angles of $\theta=0.01^{\circ}$ and $\theta=0.001^{\circ}$.
  The lowermost light red (gray) shaded cone is also for $\theta=0.001^{\circ}$, but in addition accounts exemplarily for a regeneration cavity.
  The centers of the cone-shaped areas are marked by dot-dashed lines, denoting the respective pair production thresholds, $\omega\sin\theta=2m$.
  The steep cusps at the pair creation thresholds might be smoothed and less pronounced in an actual experimental realization, see main text for details. Note that the exclusion bounds 
  for smaller angles always include the entire parameter space which can be explored for a larger angle.
  The asymptotics for $\theta\to0$ is given by the black dashed line, whereas the asymptotics for $\theta\to0$ with a 2nd cavity added is
  marked by the yellow (gray) dashed line.
  We can also provide for analytic asymptotics beyond the pair creation threshold, cf. \Eqref{eq:strong_asymp2}. Such asymptotics are shown 
  exemplarily for $\theta=0.1^{\circ}$, as a white dashed line. We stress that all these parameter regimes fulfill the condition $\frac{\epsilon eB}{\omega^2\sin^2\theta}\ll1$ for the respective angle $\theta$.
  The dotted, purple line (and area) depicts the limits \cite{Ahlers:2007qf} derived from PVLAS polarization measurements \cite{Zavattini:2007ee},
  and the blue short-dashed line (and light blue (lightest gray) area) refers to the best model-independent cosmological bounds obtained from CMB data \cite{Melchiorri:2007sq}.
  For an angle of $\theta=0.001^{\circ}$, and employing a cavity on the regeneration side, magnetically amplified tunneling with virtual minicharged particles can be expected to
  outmatch these bounds in a mass range of $2 \times 10^{-6}\lesssim m \lesssim 9 \times 10^{-5} \mathrm{eV}$.
Finally, the pink (lower-left) region shows the exclusion bounds for
$\theta=0.001^{\circ}$ attainable in a setup with second cavity and
infrared laser light at $\omega=1.165$eV, cf. footnote
\ref{fn:wavelength}.  Note that these particular bounds are derived
from the asymptotics, Eqs.~(\ref{eq:strong_asymp}) and
(\ref{eq:strong_asymp2}), without evaluating
\Eqref{eq:P_large_B_final} numerically.}
\label{fig:excl_lhs}
\end{figure}

The discovery potential of our setup can be quantified by prospective
exclusion bounds for minicharged fermions, see Figs. \ref{fig:excl_rhs} and
\ref{fig:excl_lhs}. The colored areas depict the
parameter regimes that are accessible by the proposed experimental setup, as well as different existing bounds, see below.  The
two plots concentrate on the meV mass range (Fig.~\ref{fig:excl_rhs}), and on a
mass range including very small masses (Fig.~\ref{fig:excl_lhs}). 

Fig.~\ref{fig:excl_rhs} displays the mass-coupling plane for `large'
minicharged particle masses of $m=10^{-3}\mathrm{eV}$ to $m\simeq2\mathrm{eV}$. 
This mass range is obviously irrelevant in terms of its discovery potential for
minicharged fermions, as all the regions accessible by magnetic tunneling via virtual minicharged particles
are already excluded by PVLAS \cite{Zavattini:2007ee} polarization data (cf. the lowermost, purple (gray) shaded area).
We can nevertheless obtain several interesting physical insights.
A numerical evaluation of \Eqref{eq:P_vac_final}, using $d=1\mathrm{\mu m}$, results in the blue colored area in 
Figs.~\ref{fig:excl_rhs} and~\ref{fig:excl_lhs}. 
As already discussed in \cite{Gies:2009wx}, going to
the smallest considered masses of $m=8 \times 10^{-7} \mathrm{eV}$,
the lowermost boundary of the excluded area in the absence of an external field decreases with a logarithmic dependence and finally reaches a fractional coupling of $\epsilon
\simeq 6 \times 10^{-6}$, cf. also Fig. \ref{fig:excl_lhs}.
Returning to Fig.~\ref{fig:excl_rhs}, the area above the white dashed line, is the discovery potential in the limit of weak magnetic fields for $\theta\approx0$,
cf. \Eqref{eq:P_small_B_final}.  As argued in Sect.~\ref{sec:strong}, the perturbative
calculation is only valid as long as $\frac{\epsilon e B }{m^2} \ll1$.
In Fig.~\ref{fig:excl_rhs}, the perturbative correction can be expected to be
trustworthy only to the very right of the yellow dot-dashed line, corresponding to $\frac{\epsilon e B }{m^2} =1$.
Its limit of validity is clearly exceeded before it can improve the zero-field
bounds. Therefore, the perturbative, weak-field result 
adds hardly anything to the new physics discovery potential in the zero-field limit already discussed.
For the exploration of an hitherto untested minicharged particle parameter regime, the strong-field results of Sect. \ref{sec:strong} are crucial.
A distinct feature discriminating the zero-field situation from the finite
magnetic field case is certainly the resonance in the photon-to-photon transition probability for $\omega=2m$.
For our parameters it is found at $m=1.165 \mathrm{eV}$, see Fig.~\ref{fig:excl_rhs}.

As argued in Sect. \ref{sec:strong}, for $\frac{\epsilon e B }{m^2} \gg 1$,
i.e., to the very left of the yellow dot-dashed line, the strong field approximation, \Eqref{eq:P_large_B_final}, is valid.
The additional constraint $\frac{\epsilon e B }{\omega^2\sin^2\theta}\gg 1$ is met in the entire parameter space shown in Fig.~\ref{fig:excl_rhs}.
We depict the bounds derived in the strong-field limit for two different wall thicknesses:
For $d=1\mathrm{\mu m}$, the area above the orange line, particularly the orange (light gray) area denoted by `strong', can be excluded.
Correspondingly, for $d=1\mathrm{mm}$ the area above the black dashed line can be excluded.
For minicharged particle masses below $m\simeq 5 \times 10^{-3} \mathrm{eV}$, the lowermost boundaries of these areas fall on top of each other,
and hence become independent of the wall thickness $d$.
Going to larger masses, the lines $d=1\mathrm{mm}$ and  $d=1\mathrm{\mu m}$ are found to separate, and the thicker wall always results in worse exclusion bounds 
in comparison to the thinner wall.
This constitutes a nontrivial, full numerical check of the manifestly $d$ independent asymptotics, derived analytically in \Eqref{eq:strong_asymp}.
For a given $d$, the asymptotics~(\ref{eq:strong_asymp}) are expected to become trustworthy if the condition $d\ll\frac{1}{m}\left(\frac{2m}{\omega}\right)^{-1}$ 
is met, cf. also the discussion in Sec.~\ref{sec:expparamset}.
Strikingly, this implies that the exclusion bounds to the left of the vertical dash-dotted line labeled with $d=1.8$cm, although numerically computed 
for $d=1\mathrm{\mu m}$, are also valid for wall thicknesses up to at least $d=1.8$cm, i.e., the thickness of the wall employed in the ALPS experiment.

Note that the resonance at $\omega=2m$ is clearly visible in the strong-field results in Fig.~\ref{fig:excl_rhs}.
It is, however, located in the region where  $\frac{\epsilon e B }{m^2} <1$, such that the strong field approximation cannot
be expected to hold anymore. Rather, the weak field approximation should be valid close to the resonance at $\omega=2m$. 

In Fig.~\ref{fig:excl_lhs} we present exclusion bounds in the fractional-charge mass plane ranging from minicharged particle masses of $m=8 \times
10^{-7} \mathrm{eV}$ to $m= 10^{-2} \mathrm{eV}$. The orange-colored cone-shaped area
depicts the parameter space that could be excluded by virtue of \Eqref{eq:P_large_B_final}
for an angle of $\theta=0.1^{\circ}$ and a wall thickness of
up to at least $d\lesssim1.8\mathrm{cm}$,
cf. the discussion in the context of Fig.~\ref{fig:excl_rhs} and Sec.~\ref{sec:expparamset}. 
The center of the cone-shaped area is marked by a dot-dashed line, indicating the pair production threshold $\omega\sin\theta=2m$ 
for the respective incidence angle (cf. Sec.~\ref{sec:strong}).

Following the orange-colored exclusion bounds for
$\theta=0.1^{\circ}$, the cone-shaped areas in pink and dark red
(from right to left) correspond to the explorable parameter space at
$\theta=0.01^{\circ}$ and $\theta=0.001^{\circ}$. The lowermost
light-red cone shaped area corresponds to $\theta=0.001^{\circ}$ and
the additional use of a second cavity.  For completeness, the
exclusion bounds for $\theta=0.001^{\circ}$ attainable in a setup with
second cavity and infrared laser light at $\omega=1.165$eV are
represented by the pink (lower-left) region. This might constitute a more favorable choice for the laser energy in a
setting with regeneration cavity, cf. also footnote
\ref{fn:wavelength}. For simplicity, we have
estimated this region from the asymptotic formulas
(\ref{eq:strong_asymp}) and (\ref{eq:strong_asymp2}).  Note that in this situation also the resonance
corresponding to $\theta=0.001^{\circ}$ is shifted according to
\Eqref{eq:pair}.  Remarkably, for this choice of parameters, the
bounds derived from cosmological observations can be overcome in an
even larger mass range.

The lowermost boundary of the accessible parameter regime in the asymptotic limit $\theta\to0$ is depicted by the black dashed line.
The yellow dashed line shows the same asymptotics, but in addition assumes resonant regeneration with a second cavity installed behind
the wall, employing ${\cal N}=10^5$. As the transition probability
$P_{\gamma\rightarrow \gamma}^{(\mathrm{strong})} \sim \epsilon^6$,
cf. \Eqref{eq:P_large_B_final}, employing the second cavity yields almost an
order of magnitude better exclusion bounds.
Note that in order to achieve the same effect by tuning the magnetic
field strength, one would correspondingly need to enhance the magnetic
field by a factor of $\sqrt{10^5}$ as it enters \Eqref{eq:P_large_B_final} by its square.
This would demand for magnetic fields of $B\sim 1.5 \times 10^3 \mathrm T$,
corresponding to field strengths that can only be obtained in highly focused
lasers in a laboratory. However, there of course the extent of the field
itself would be limited to a smaller region and requires to go beyond the homogeneous field approximation used here.  Thus, a large effort on the 
optics side seems more beneficial than requiring increasingly stronger magnetic fields.

To allow for a comparison of our results with current experimental limits, Fig.~\ref{fig:excl_lhs} includes the exclusion limits
\cite{Ahlers:2007qf} as provided by PVLAS polarization measurements \cite{Zavattini:2007ee} (upper dotted purple line and area).
The light blue area, limited from below by the blue dotted line, depicts the best model-independent cosmological bounds obtained from CMB data
\cite{Melchiorri:2007sq}.

To summarize: even with conservatively chosen experimental parameters, we
find that the LSW scenario with virtual minicharged fermionic
particle-antiparticle states could improve the limits provided by the PVLAS collaboration below
$m\lesssim 10^{-4} \mathrm{eV}$, and even outmatch cosmological bounds derived
from CMB data below $m \lesssim 5 \times 10^{-6} \mathrm{eV}$ for extremely small angles $\theta\to0$.  Employing a
cavity on the regeneration side, these values can be even improved to outmatch
PVLAS bounds for $m\lesssim 2 \times 10^{-3} \mathrm{eV}$ and to overcome cosmological
bounds at $m \lesssim 9 \times 10^{-5} \mathrm{eV}$.

\section{Conclusions \label{sec:conclusion}}

In this paper, we have generalized the recently proposed light-shining-through-walls 
scenario via virtual minicharged particles \cite{Gies:2009wx} to account for a homogeneous external magnetic field. 
Moreover, we have demonstrated how this particular LSW scenario can be employed to constrain the parameter space for minicharged fermions in a dedicated laboratory experiment.
Whereas in the absence of an external field, the tunneling scenario with virtual minicharged particles has only little sensitivity to a new-physics parameter space \cite{Gies:2009wx},
we have shown that an external magnetic field can significantly enhance its discovery potential.
Notably, magnetically amplified `tunneling of the 3rd kind' even allows, in a rather small-scale laboratory setup, to explore a parameter regime which has been inaccessible so far.

To this end, we have suggested a simple experimental setting, involving only present-day technology.
In contrast to standard LSW setups, typically employing $\theta=\varangle({\bf B},{\bf k})=\pi/2$,
our experiment requires the probe photons to enter at a preferably very small angle $\theta$ with respect to the homogeneous magnetic field lines.
Together with a resonant regeneration cavity, it
has the potential to outmatch the currently best direct laboratory bounds on minicharged particles in the sub-$\mathrm{m eV}$
mass range, as provided by the PVLAS collaboration.
In the range of minicharged particle masses below $ \mathcal{O} (10^{-4})\mathrm{eV}$, it could explore -- and even overcome -- a parameter regime so far only accessible in 
large scale cosmological observations.

We note that similar ideas could also be considered on terrestrial or astrophysical scales, 
as magnetic fields of larger extent might provide access to extended regions of the minicharged particle parameter space. 
Also, a modification of this setup beyond the optical regime might be favorable in the search of higher-mass minicharged particle dark matter \cite{Cheung:2007ut,Feldman:2007wj}.

From a formal point of view, our analysis substantially builds on analytic insights into the QED photon
polarization tensor at one loop accuracy, while retaining its full momentum dependence.
Within standard LSW scenarios, the barrier is traversed by means of real particles, whose energy and momentum is fixed by energy-momentum conservation.
By contrast, the tunneling scenario via virtual particles genuinely requires to account for the entire range of allowed four momenta within the particle-antiparticle loop.
This significantly complicates the corresponding calculation, but also gives rise to several novel features, as compared to conventional LSW scenarios.

We have derived analytical expressions for the photon-to-photon transition probability in the limiting cases of both, weak and strong magnetic fields,
corresponding to $\left\{\frac{\epsilon e B }{m^2},\frac{\epsilon e B }{\omega^2}\right\} \ll 1$ and $\left\{\frac{\epsilon e B }{m^2},\frac{\epsilon e B }{\omega^2\sin^2\theta}\right\}  
\gg 1$, respectively.
An exemplary numerical evaluation of these expressions, using experimental parameters which can be feasibly implemented, has allowed us to provide exclusion bounds for fermionic minicharged 
particles in a wide range of the fractional-charge mass plane.

In addition, we have analytically extracted simple asymptotics for the photon-to-photon transition probability in the most relevant limiting cases, such that the obtained 
results can be straightforwardly adopted to different experimental parameter sets.   

In summary, the relatively small experimental effort required to realize the ``Tunneling of Light via Loops'' experiment proposed by us compares favorably with the prospect of exploring a parameter
space for minicharged particles hitherto inaccessible in the laboratory. An implementation of this setup thus seems 
highly worthwhile and will put the existence of minicharged particles further to the test.

\section*{Acknowledgments}

BD, HG and NN acknowledge support by the DFG under grants SFB-TR18 and
GI~328/5-2 (Heisenberg program) as well as GRK-1523.  We thank J.~Jaeckel, S.~Keppler, A.~Lindner, J.~Pauli, D.~Trines and P.~Winkler
for interesting discussions and helpful correspondence.

\appendix

\section{Scalar minicharged particles \label{sec:mini_bosons}}

A perturbative expansion of the photon polarization tensor for scalar minicharged particles \cite{Schubert:2000yt} analogous to \Eqref{eq:PI_pert} yields
\begin{align}
 \left\{
 \begin{array}{c}
\Pi^{\rm sc}_1\\
 \Pi^{\rm sc}_2\\
 \Pi^{\rm sc}_3
 \end{array}
\right\}=\Pi^{{\rm sc}(0)}+
\left\{
 \begin{array}{c}
\Pi^{{\rm sc}(2)}_1\\ 
\Pi^{{\rm sc}(2)}_2\\
 \Pi^{{\rm sc}(2)}_3
 \end{array}
\right\}+{\cal O}\left((\epsilon eB)^4\right).
\label{eq:PI_pert_scalar}
\end{align}
The zero-field expression reads
\begin{equation}
\label{eq:pol_vac_scalar}
 \Pi^{{\rm sc}(0)}=-(k^2)^2\,\frac{\epsilon^2\alpha}{24\pi}\int\limits_0^1{\rm d}\nu\,\frac{\nu^4}{\Phi_0}\ ,
\end{equation}
and the contribution at ${\cal O}\left((\epsilon eB)^2\right)$ is given by
\begin{align}
&\hspace*{-3mm}\left\{\!
 \begin{array}{c}
\Pi^{{\rm sc}(2)}_1\\ 
 \Pi^{{\rm sc}(2)}_2\\
 \Pi^{{\rm sc}(2)}_3
 \end{array}
\!\right\}
=-\frac{\epsilon^2\alpha (\epsilon eB)^2}{24\pi}\int\limits_{0}^{1}{\rm d}\nu\,\frac{\nu^2(2-\nu^2)}{\Phi_0^2} \nonumber\\
&\hspace*{-3mm}\times\!\left[\!
\left\{\!\!\!
\begin{array}{c}
\frac{1}{2-\nu^2}\\
1\\
1
 \end{array}
\!\!\!\right\}
k_{\parallel}^2
+\!
\left(\left\{\!\!\!
\begin{array}{c}
1\\
\frac{3-\nu^4}{2\nu^2(2-\nu^2)}\\
1
\end{array}
\!\!\!\right\}-\frac{\frac{(1-\nu^2)^2}{2-\nu^2}}{\Phi_0}\frac{k^2}{4}\!\right)\!k_{\perp}^2
\right]\!,
\label{eq:PI_pert2_scalar}
\end{align}
with $\Phi_0$ defined in \Eqref{eq:phi0}.
Implementing the limit $\epsilon eB\to\infty$ in the polarization tensor for scalar minicharged particles, the first nonvanishing contributions are at most logarithmic in $\epsilon eB$.
Hence, in comparison to the photon polarization tensor for fermionic minicharged particles in the strong-field limit, whose leading contribution is $\sim\epsilon eB$ [cf. \Eqref{eq:PI_strong}],
the photon polarization tensor for scalar minicharged particles in the strong-field limit is suppressed and starts to contribute only at subleading order \cite{DoebrichKarbstein}.
The reason for this qualitative difference lies in the fact that the bosonic fluctuation spectrum in
the presence of a magnetic field does not exhibit a near-zero-mode. We conclude that a Pauli-type spin-field coupling is essential for the physics discussed in the
main part of this work. Hence, given the same experimental parameters, we expect the exclusion bounds on scalar minicharged
particles, achievable via `tunneling of the 3rd kind' to be much weaker.

\section{The regime of validity of the analytic asymptotics in the small-mass limit \label{sec:omega*d}}

In this appendix, we want to comment on the range of validity of the various analytic asymptotics in the small-mass limit, $\omega\gg2m$, derived in the main text.
The asymptotics turned out to be independent of the thickness $d$ of the barrier. Here we study the $d$ regime where 
these asymptotics can be considered as trustworthy.

To derive the asymptotics, we basically expanded the integrand in $f_>$ around $\lambda=1$, and performed the $\lambda$ integration thereafter.
We thereby  also approximated  in particular the exponential part in the integrand as
\begin{equation}
 e^{i\omega d\lambda}\simeq e^{i\omega d}.
\label{eq:exp_lambda=1}
\end{equation}
The latter term, being a pure phase factor, canceled in the determination of the photon-to-photon transition probability, as it involves an overall modulus-squared, cf. \Eqref{eq:def_f}.
Obviously the approximation in \Eqref{eq:exp_lambda=1} is well-justified for $\omega d\ll 1$, as to leading order in $\omega d\ll 1$, both exponentials in \Eqref{eq:exp_lambda=1} are 
equal to one.
Its regime of validity in the opposite limit $\omega d\gg1$ is less clear. Hence, we subsequently focus on the limit $\omega d\gg1$.
In order to estimate when this approximation is justified, we study the following integral,
\begin{align}
 \int_0^{\sqrt{1-\frac{4m^2}{\omega^2}}}{\rm d}\lambda\,\frac{e^{i\omega d\lambda}}{(1-\lambda)^{n/2}}\,,
\label{eq:posterint}
\end{align}
for $n=3,5,7,\ldots\,$, both by setting $\lambda=1$ in the exponential from the outset, and by explicitly performing the integration over $\lambda$.
Note, that \Eqref{eq:posterint} exhibits the basic features of the integrals encountered in deriving the asymptotics, and can be integrated explicitly.
Setting $\lambda=1$ in the exponential, we obtain
\begin{align}
&\int_0^{\sqrt{1-\frac{4m^2}{\omega^2}}}{\rm d}\lambda\,\frac{e^{i\omega d}}{(1-\lambda)^{n/2}} \nonumber\\
&=\frac{2}{n-2}\,e^{i\omega d}\left[\left(1-\sqrt{1-\tfrac{4m^2}{\omega^2}}\right)^{\tfrac{2-n}{2}}-1\right] \nonumber\\
&=e^{i\omega d}\,\frac{\left(\!\sqrt{2}\right)^{n}}{n-2}\left(\frac{1}{\frac{2m}{\omega}}\right)^{n-2}\left[1+{\cal O}\left(\left(\tfrac{2m}{\omega}\right)^2\right)\right],
\label{eq:app_l=1}
\end{align}
whereas in the other case, we first reformulate the corresponding expression with the help of the substitution $\lambda\to 1-l$. This results is
\begin{align}
 &\int_0^{\sqrt{1-\frac{4m^2}{\omega^2}}}{\rm d}\lambda\,\frac{e^{i\omega d\lambda}}{(1-\lambda)^{n/2}}
 =e^{i\omega d}\int^1_{1-\sqrt{1-\frac{4m^2}{\omega^2}}}{\rm d}l\,\frac{e^{-i\omega dl}}{l^{n/2}} \nonumber\\
 &=e^{i\omega d}\int^1_{\frac{1}{2}\left(\frac{2m}{\omega}\right)^2+{\cal O}\left(\left(\frac{2m}{\omega}\right)^4\right)}{\rm d}l\,\frac{e^{-i\omega dl}}{l^{n/2}}.
\label{eq:app_l}
\end{align}
We now focus on the evaluation of the integral in \Eqref{eq:app_l}.
For $n$ integer and odd, we obtain
 \begin{multline}
 \int {\rm d}l\ \frac{e^{-i\omega dl}}{l^{n/2}}=\frac {i}{\omega d} \biggl\{{\frac{{e}^{-i\omega dl}}{{l}^{n/2}}}  \\
+ \frac{n}{2}\left(i\omega d \right) ^{n/2}\left[ \Gamma 
 \left( -\tfrac{n}{2} \right) -\Gamma  \left( -\tfrac{n}{2},i\omega d l \right)  \right]   \biggr\} +C\,,
 \label{eq:intsapprox}
\end{multline}
with integration constant $C$.
The incomplete gamma function $\Gamma(\alpha,\chi)$ has an exact series representation (formula 8.354.2 of~\cite{Gradshteyn}),
\begin{equation}
 \Gamma(\alpha,\chi)=\Gamma(\alpha)-\sum_{k=0}^\infty\frac{(-1)^k\chi^{\alpha+k}}{k!\,(\alpha+k)}\ ,\ \ \alpha\neq-\mathbb{N}_0\ ,
\label{eq:grad1}
\end{equation}
and its asymptotic expansion for $|\chi|\to\infty$ is given by (formula 8.357 of~\cite{Gradshteyn})
\begin{equation}
 \Gamma(\alpha,\chi)=e^{-\chi}\,\chi^{\alpha-1}\left[1+{\cal O}\left(\tfrac{1}{|\chi|}\right)\right].
\label{eq:grad2}
\end{equation}
with the additional requirement $-\tfrac{3\pi}{2}<{\rm arg}\,\chi<\tfrac{3\pi}{2}$.
Equation~(\ref{eq:grad1}) implies
\begin{multline}
 \frac{i}{\omega d}\left\{{\frac{{e}^{-i\omega dl}}{{l}^{n/2}}}+\frac{n}{2}\left(i\omega d \right) ^{n/2}\left[ \Gamma\left(-\tfrac{n}{2}\right)-\Gamma\left(-\tfrac{n}{2},i\omega d l\right) \right]\right\} \\
=\frac{l}{l^{n/2}}\sum_{k=0}^\infty\frac{(-i)^{k}(\omega dl)^{k}}{k!\,\left(k+1-\tfrac{n}{2}\right)}\ .
\end{multline}
We use this expression to evaluate the integral in \Eqref{eq:app_l} at the lower integration limit.
As we are interested in the limit $\omega d\gg1$ here, at the upper integration limit we rather employ \Eqref{eq:grad2}, yielding
\begin{equation}
 \left(i\omega d \right) ^{n/2}\Gamma\left(-\tfrac{n}{2},i\omega d\right)=-i\,\frac{e^{-i\omega d}}{\omega d}\left[1+{\cal O}\left(\tfrac{1}{\omega d }\right)\right].
\end{equation}
Hence, \Eqref{eq:app_l} can be cast in the following form
\begin{align}
 &\int_0^{\sqrt{1-\frac{4m^2}{\omega^2}}}{\rm d}\lambda\,\frac{e^{i\omega d\lambda}}{(1-\lambda)^{n/2}} \nonumber\\
 =&e^{i\omega d}\Biggl\{\left(\!\sqrt{i}\right)^{n+2}\frac{n}{2}\,\Gamma \left( -\tfrac{n}{2} \right)\left(\!\sqrt{\omega d}\right)^{n-2} +{\cal O}\left(\tfrac{1}{\omega d}\right) \nonumber\\
 &\quad\quad+\left(\!\sqrt{2}\right)^{n}\!\left(\frac{1}{\frac{2m}{\omega}}\right)^{n-2}\sum_{k=0}^\infty
 \frac{(-i)^{k}\left(\frac{2m}{\omega}dm\right)^k}{k!\,\left(n-2-2k\right)} \nonumber\\
 &\quad\quad\quad\quad\times\left[1+{\cal O}\left(\left(\tfrac{2m}{\omega}\right)^2\right)\right]\Biggr\}\,.
\label{eq:app_l2}
\end{align}
The $k=0$ term in the sum in \Eqref{eq:app_l2} agrees with the last line of \Eqref{eq:app_l=1}. For $n=3,5,7,\ldots\,$, and given that
\begin{equation}
 \frac{2m}{\omega}dm\ll1 \quad\quad\leftrightarrow\quad\quad d\ll\frac{1}{m}\left(\frac{2m}{\omega}\right)^{-1},
\label{eq:condd}
\end{equation}
it is the leading contribution to \Eqref{eq:app_l2}. If \Eqref{eq:condd} is met, both terms due to $k\geq1$ in the sum, and the term in the second line of \Eqref{eq:app_l2}, become negligible as compared to the $k=0$ term. 
In this limit, Eqs.~(\ref{eq:app_l=1}) and (\ref{eq:app_l2}) indeed coincide.
Hence, the various asymptotics derived for $\omega\gg 2m$ in the main text should be trustworthy for thicknesses $d$ of the barrier fulfilling the condition $d\ll\frac{1}{m}\left(\frac{2m}{\omega}\right)^{-1}$. 

\section{$f_>^{({\rm strong})}$ beyond the pair creation threshold \label{app:lambda}}

As pointed out in Sec.~\ref{sec:towards_fs}, the pole at $\lambda=\cos\theta-i\eta$ becomes relevant in any explicit evaluation of the $\lambda$ contribution,
if the condition in \Eqref{eq:polcond} is met. It is then convenient to rewrite the integral in \Eqref{eq:lambda_large_B} in a manifestly $\eta$ independent way. 
This can be achieved, e.g., with the help of an integration by parts. Thereafter, we obtain the following rather lengthy, but $\eta$ independent and finite expression,
\begin{widetext}
 \begin{align}
&\hspace*{-5mm}\int_0^{\sqrt{1-\tfrac{4m^2}{\omega^2}}}{\rm d}\lambda\ \frac{\left(1-\lambda^2-\tfrac{4m^2}{\omega^2}\right)^{-1/2}}{(\cos\theta-{i}\eta-\lambda)^2}\,\frac{{e}^{{i}\frac{\omega d}{\cos\theta}\lambda}}{\sqrt{1-\lambda^2}(\lambda+\cos\theta)}
=\left(1-\tfrac{\omega\cos\theta}{2m}\ \tfrac{\exp\left(i\frac{\omega d}{\cos\theta}\sqrt{1-\tfrac{4m^2}{\omega^2}}\right)}{\sqrt{1-\tfrac{4m^2}{\omega^2}}+\cos\theta}\right)\tfrac{\ln\left(2\sqrt{1-\tfrac{4m^2}{\omega^2}}\right)}{\left(\sin^2\theta-\tfrac{4m^2}{\omega^2}\right)^{3/2}} \nonumber\\
&+\tfrac{g(\cos\theta)\,\sqrt{1-\tfrac{4m^2}{\omega^2}}\left[\cos\theta-\sqrt{1-\tfrac{4m^2}{\omega^2}}\,\ln(\cos\theta)  + \left(\sqrt{1-\tfrac{4m^2}{\omega^2}}-\cos\theta\right)\ln\left(\sqrt{1-\tfrac{4m^2}{\omega^2}}-\cos\theta\right)\right]
-\ln(\cos\theta)-i\pi+\ln\!\left(\sqrt{\sin^2\theta-\tfrac{4m^2}{\omega^2}}+\sqrt{1-\tfrac{4m^2}{\omega^2}}\right)}{\left(\sin^2\theta-\tfrac{4m^2}{\omega^2}\right)^{3/2}} \nonumber\\
&-\tfrac{\sqrt{1-\tfrac{4m^2}{\omega^2}}}{\left(\sin^2\theta-\tfrac{4m^2}{\omega^2}\right)\cos^2\theta}-\tfrac{{i}\pi g(\cos\theta)}{\sqrt{\sin^2\theta-\tfrac{4m^2}{\omega^2}}}
-\tfrac{{i}\pi\cos\theta}{\left(\sin^2\theta-\tfrac{4m^2}{\omega^2}\right)^{3/2}}\int_0^{\cos\theta}{\rm d}\lambda\ g(\lambda) \ +\ \tfrac{\ln\left[2\left(1-\tfrac{4m^2}{\omega^2}\right)\right]\cos\theta}{\left(\sin^2\theta-\tfrac{4m^2}{\omega^2}\right)^{3/2}}\int_0^{\sqrt{1-\tfrac{4m^2}{\omega^2}}}{\rm d}\lambda\ g(\lambda) \nonumber\\
&+\int_0^{\sqrt{1-\tfrac{4m^2}{\omega^2}}}{\rm d}\lambda\ \tfrac{g(\lambda)\sqrt{1-\tfrac{4m^2}{\omega^2}-\lambda^2}-g(\cos\theta)\sqrt{\sin^2\theta-\tfrac{4m^2}{\omega^2}}}{\left(\sin^2\theta-\tfrac{4m^2}{\omega^2}\right)\left(\lambda-\cos\theta\right)}
\ +\ \cos\theta \int_{0}^{\sqrt{1-\tfrac{4m^2}{\omega^2}}}{\rm d}\lambda\ \tfrac{\left[g(\cos\theta)-g(\lambda)\right]\ \ln\left|\cos\theta-\lambda\right|}{\left(\sin^2\theta-\tfrac{4m^2}{\omega^2}\right)^{3/2}} \nonumber\\
&+\tfrac{\cos\theta}{\left(\sin^2\theta-\tfrac{4m^2}{\omega^2}\right)^{3/2}}\int_0^{\sqrt{1-\tfrac{4m^2}{\omega^2}}}{\rm d}\lambda\ g(\lambda)\ 
\ln\!\left(1-\tfrac{\lambda\cos\theta-\sqrt{\left(1-\tfrac{4m^2}{\omega^2}-\lambda^2\right)\!\left(\sin^2\theta-\tfrac{4m^2}{\omega^2}\right)}}{1-\tfrac{4m^2}{\omega^2}}\right),
\label{eq:superlangda}
\end{align}
\end{widetext}
where we introduced
\begin{align}
 g(\lambda)=\frac{{e}^{{i}\frac{\omega d}{\cos\theta}\lambda}\left(\frac{\lambda}{1-\lambda^2}-\frac{1}{\lambda+\cos\theta}+\frac{i \omega d}{\cos\theta}\right)}{\sqrt{1-\lambda^2}(\lambda+\cos\theta)}\,. \label{fct1}
\end{align}
Equation~(\ref{eq:superlangda}) is not only essential for the purpose of explicit numerical calculations in the regime~(\ref{eq:polcond}), it also allows for the extraction of an analytical asymptotics for the photon-to-photon transition probability in the limit $\omega \gg2 m$ and $d\ll\frac{1}{m}\left(\frac{2m}{\omega}\right)^{-1}$ (cf. App.~\ref{sec:omega*d}).

An expansion in $\frac{2m}{\omega}\ll1$ of the terms in \Eqref{eq:superlangda} featuring no integrations over the parameter $\lambda$ results in
\begin{equation}
 c_1=-\frac{e^{i\frac{\omega d}{\cos\theta}}\,\cos\theta\,\ln2}{(1+\cos\theta)\sin^3\theta}\ \frac{1}{\frac{2m}{\omega}}+{\cal O}\left(\left(\tfrac{2m}{\omega}\right)^0\right).
\end{equation}
We now consider the contributions to \Eqref{eq:superlangda} involving integrations over $\lambda$. The first integral yields
\begin{equation}
 c_2=\tfrac{-{i}\pi\cos\theta}{\left(\sin^2\theta-\tfrac{4m^2}{\omega^2}\right)^{3/2}}\int\limits_0^{\cos\theta}{\rm d}\lambda\ g(\lambda)={\cal O}\left(\left(\tfrac{2m}{\omega}\right)^0\right),
\end{equation}
and (cf. the discussion in the main text),
\begin{align}
 c_3&=\tfrac{\ln\left[2\left(1-\tfrac{4m^2}{\omega^2}\right)\right]\cos\theta}{\left(\sin^2\theta-\tfrac{4m^2}{\omega^2}\right)^{3/2}}\int_0^{\sqrt{1-\tfrac{4m^2}{\omega^2}}}{\rm d}\lambda\ g(\lambda) \nonumber\\
    &\approx\tfrac{\ln\left[2\left(1-\tfrac{4m^2}{\omega^2}\right)\right]\tfrac{\cos\theta}{1+\cos\theta}}{2\sqrt{2}\left(\sin^2\theta-\tfrac{4m^2}{\omega^2}\right)^{3/2}}\int_0^{\sqrt{1-\tfrac{4m^2}{\omega^2}}}{\rm d}\lambda\ \frac{e^{i\frac{\omega d}{\cos\theta}}}{\left(1-\lambda\right)^{3/2}} \nonumber\\
    &=\frac{e^{i\frac{\omega d}{\cos\theta}}\,\cos\theta\,\ln2}{(1+\cos\theta)\sin^3\theta}\ \frac{1}{\frac{2m}{\omega}}+{\cal O}\left(\left(\tfrac{2m}{\omega}\right)^0\right).
\end{align}
Moreover, we obtain
\begin{align}
 c_4&=\int_0^{\sqrt{1-\tfrac{4m^2}{\omega^2}}}{\rm d}\lambda\ \tfrac{g(\lambda)\sqrt{1-\tfrac{4m^2}{\omega^2}-\lambda^2}-g(\cos\theta)\sqrt{\sin^2\theta-\tfrac{4m^2}{\omega^2}}}{\left(\sin^2\theta-\tfrac{4m^2}{\omega^2}\right)\left(\lambda-\cos\theta\right)} \nonumber\\
    &=\int_0^{\sqrt{1-\tfrac{4m^2}{\omega^2}}}{\rm d}\lambda\ \tfrac{g(\lambda)\sqrt{1-\tfrac{4m^2}{\omega^2}-\lambda^2}}{\left(\sin^2\theta-\tfrac{4m^2}{\omega^2}\right)\left(\lambda-\cos\theta\right)}
+{\cal O}\left(\left(\tfrac{2m}{\omega}\right)^0\right) \nonumber\\
    &\approx\tfrac{e^{i\frac{\omega d}{\cos\theta}}(1+\cos\theta)^{-1}}{2\left(\sin^2\theta-\tfrac{4m^2}{\omega^2}\right)}\int_0^{\sqrt{1-\tfrac{4m^2}{\omega^2}}}{\rm d}\lambda\ \tfrac{\left(\sqrt{1-\tfrac{4m^2}{\omega^2}}-\lambda\right)^{1/2}}{\left(\lambda-\cos\theta\right)\left(1-\lambda\right)^{3/2}} \nonumber\\
&\hspace*{4cm}+{\cal O}\left(\left(\tfrac{2m}{\omega}\right)^0\right) \nonumber\\
&=-\frac{e^{i\frac{\omega d}{\cos\theta}}}{\sin^4\theta}\,\ln\left(\frac{2m}{\omega}\right)+{\cal O}\left(\left(\tfrac{2m}{\omega}\right)^0\right),
\end{align}
and
\begin{align}
 c_5&=\cos\theta \int_{0}^{\sqrt{1-\tfrac{4m^2}{\omega^2}}}{\rm d}\lambda\ \tfrac{\left[g(\cos\theta)-g(\lambda)\right]\ \ln\left|\cos\theta-\lambda\right|}{\left(\sin^2\theta-\tfrac{4m^2}{\omega^2}\right)^{3/2}} \nonumber\\
    &=\cos\theta \int_{0}^{\sqrt{1-\tfrac{4m^2}{\omega^2}}}{\rm d}\lambda\ \tfrac{-g(\lambda)\ \ln\left|\cos\theta-\lambda\right|}{\left(\sin^2\theta-\tfrac{4m^2}{\omega^2}\right)^{3/2}}+{\cal O}\left(\left(\tfrac{2m}{\omega}\right)^0\right) \nonumber\\
    &\approx\tfrac{-\tfrac{\cos\theta\ e^{i\frac{\omega d}{\cos\theta}}}{2\sqrt{2}(1+\cos\theta)}}{\left(\!\sin^2\theta-\tfrac{4m^2}{\omega^2}\!\right)^{3/2}}\!\int_{0}^{\sqrt{1-\tfrac{4m^2}{\omega^2}}}\!\!{\rm d}\lambda\,\tfrac{\ln\left|\cos\theta-\lambda\right|}{(1-\lambda)^{3/2}} +{\cal O}\!\left(\left(\tfrac{2m}{\omega}\right)^0\right) \nonumber\\
    &=-\frac{e^{i\frac{\omega d}{\cos\theta}}\cos\theta\,\ln(1-\cos\theta)}{(1+\cos\theta)\sin^3\theta}\,\frac{1}{\frac{2m}{\omega}}+{\cal O}\!\left(\left(\tfrac{2m}{\omega}\right)^0\right)\!,
\end{align}
\pagebreak
as well as
\begin{align}
 c_6&=\tfrac{\cos\theta}{\left(\sin^2\theta-\tfrac{4m^2}{\omega^2}\right)^{3/2}}\int_0^{\sqrt{1-\tfrac{4m^2}{\omega^2}}}{\rm d}\lambda\ g(\lambda) \nonumber\\
&\quad\times\ln\left(1-\tfrac{\lambda\cos\theta-\sqrt{\left(1-\tfrac{4m^2}{\omega^2}-\lambda^2\right)\!\left(\sin^2\theta-\tfrac{4m^2}{\omega^2}\right)}}{1-\tfrac{4m^2}{\omega^2}}\right) \nonumber\\
&\approx\tfrac{\frac{\cos\theta}{1+\cos\theta}e^{i\frac{\omega d}{\cos\theta}}}{2\sqrt{2}\left(\sin^2\theta-\tfrac{4m^2}{\omega^2}\right)^{3/2}}\int_0^{\sqrt{1-\tfrac{4m^2}{\omega^2}}}{\rm d}\lambda\ \frac{1}{\left(1-\lambda\right)^{3/2}} \nonumber\\
&\,\times\ln\!\left(\!1-\tfrac{\sqrt{1-\frac{4m^2}{\omega^2}}\cos\theta-\sqrt{2\left(\sqrt{1-\tfrac{4m^2}{\omega^2}}-\lambda\right)\!\left(\!\sin^2\theta-\tfrac{4m^2}{\omega^2}\!\right)}}{1-\tfrac{4m^2}{\omega^2}}\right) \nonumber\\
&=\frac{e^{i\frac{\omega d}{\cos\theta}}\,\cos\theta\,\ln(1-\cos\theta)}{(1+\cos\theta)\sin^3\theta}\ \frac{1}{\frac{2m}{\omega}} \nonumber\\
&\hspace*{1cm}-\frac{e^{i\frac{\omega d}{\cos\theta}}\,\cos\theta}{\sin^4\theta}\ln\left(\frac{2m}{\omega}\right)+{\cal O}\left(\left(\tfrac{2m}{\omega}\right)^0\right).
\end{align}
Hence, the leading behavior of \Eqref{eq:superlangda} in the limit $\frac{2m}{\omega}\ll1$ is given by
\begin{equation}
 \sum_{j=1}^6 c_j=-\tfrac{(1+\cos\theta) e^{i\frac{\omega d}{\cos\theta}}}{\sin^4\theta}\,\ln\left(\tfrac{2m}{\omega}\right)+{\cal O}\left(\left(\tfrac{2m}{\omega}\right)^0\right).
\end{equation}
The terms $\sim(\frac{2m}{\omega})^{-1}$, encountered in some of the contributions $c_j$ cancel, resulting in an overall logarithmic dependence on the ratio $\frac{2m}{\omega}$.


\begin{thebibliography}{999}

\bibitem{Gies:2007ua}
  H.~Gies,
  J.\ Phys.\ A  {\bf 41}, 164039 (2008)
  [arXiv:0711.1337 [hep-ph]]. 

\bibitem{Jaeckel:2010ni}
  J.~Jaeckel, A.~Ringwald,
  Ann.\ Rev.\ Nucl.\ Part.\ Sci.\  {\bf 60}, 405-437 (2010).
  [arXiv:1002.0329 [hep-ph]].

\bibitem{Cheung:2007ut} 
  K.~Cheung and T.~-C.~Yuan,
  JHEP {\bf 0703}, 120 (2007)
  [hep-ph/0701107].

\bibitem{Feldman:2007wj} 
  D.~Feldman, Z.~Liu and P.~Nath,
  Phys.\ Rev.\ D {\bf 75}, 115001 (2007)
  [hep-ph/0702123 [HEP-PH]].

\bibitem{Arias:2012mb} 
  P.~Arias, D.~Cadamuro, M.~Goodsell, J.~Jaeckel, J.~Redondo and A.~Ringwald,
  arXiv:1201.5902 [hep-ph].

\bibitem{Cline:2012is} 
  J.~M.~Cline, Z.~Liu and W.~Xue,
  Phys.\ Rev.\ D {\bf 85}, 101302 (2012)
  [arXiv:1201.4858 [hep-ph]].

\bibitem{Chou:2007zzc}
  A.~S.~Chou {\it et al.}  [GammeV (T-969) Collaboration],
  Phys.\ Rev.\ Lett.\  {\bf 100}, 080402 (2008)
  [arXiv:0710.3783 [hep-ex]].

\bibitem{Robilliard:2007bq}
  C.~Robilliard, R.~Battesti, M.~Fouche, J.~Mauchain, A.~M.~Sautivet, F.~Amiranoff and C.~Rizzo,
  Phys.\ Rev.\ Lett.\  {\bf 99}, 190403 (2007)
  [arXiv:0707.1296 [hep-ex]].

\bibitem{Fouche:2008jk}
  M.~Fouche {\it et al.},
  Phys.\ Rev.\  D {\bf 78}, 032013 (2008)
  [arXiv:0808.2800 [hep-ex]].

\bibitem{Afanasev:2008jt}
  A.~Afanasev {\it et al.},
  Phys.\ Rev.\ Lett.\  {\bf 101}, 120401 (2008)
  [arXiv:0806.2631 [hep-ex]].

\bibitem{Pugnat:2007nu}
  P.~Pugnat {\it et al.}  [OSQAR Collaboration],
  Phys.\ Rev.\  D {\bf 78}, 092003 (2008)
  [arXiv:0712.3362 [hep-ex]].

\bibitem{Battesti:2010dm}
  R.~Battesti, M.~Fouche, C.~Detlefs, T.~Roth, P.~Berceau, F.~Duc, P.~Frings, G.~L.~J.~A.~Rikken {\it et al.},
  Phys.\ Rev.\ Lett.\  {\bf 105}, 250405 (2010).
  [arXiv:1008.2672 [hep-ex]].


\bibitem{Ehret:2010mh}
  K.~Ehret, M.~Frede, S.~Ghazaryan {\it et al.},
  Phys.\ Lett.\  {\bf B689}, 149-155 (2010).
  [arXiv:1004.1313 [hep-ex]].

\bibitem{Okun:1982xi}
  L.~B.~Okun,
  Sov.\ Phys.\ JETP {\bf 56}, 502 (1982)
  [Zh.\ Eksp.\ Teor.\ Fiz.\  {\bf 83}, 892 (1982)].

\bibitem{Holdom:1985ag}
  B.~Holdom,
  Phys.\ Lett.\  B {\bf 166}, 196 (1986).

\bibitem{Anselm:1986gz}
  A.~A.~Anselm,
  Yad.\ Fiz.\  {\bf 42}, 1480 (1985). 

\bibitem{Gies:2006ca}
  H.~Gies, J.~Jaeckel and A.~Ringwald,
  Phys.\ Rev.\ Lett.\  {\bf 97}, 140402 (2006)
  [arXiv:hep-ph/0607118].

\bibitem{Redondo:2010dp} 
  J.~Redondo and A.~Ringwald,
  Contemp.\ Phys.\  {\bf 52}, 211 (2011)
  [arXiv:1011.3741 [hep-ph]].

\bibitem{Sikivie:1983ip}
  P.~Sikivie,
  Phys.\ Rev.\ Lett.\  {\bf 51} (1983) 1415
  [Erratum-ibid.\  {\bf 52} (1984) 695].

\bibitem{VanBibber:1987rq}
K.~Van~Bibber {\it et al.}
\newblock Phys. Rev. Lett. {\bf 59}, 759 (1987).

\bibitem{Steffen:2009sc}
  J.~H.~Steffen and A.~Upadhye,
  Mod.\ Phys.\ Lett.\  A {\bf 24}, 2053 (2009)
  [arXiv:0908.1529 [hep-ex]].

\bibitem{Maiani:1986md}
  L.~Maiani, R.~Petronzio and E.~Zavattini,
  Phys.\ Lett.\ B {\bf 175}, 359 (1986). 

\bibitem{Raffelt:1987im}
  G.~Raffelt and L.~Stodolsky,
  Phys.\ Rev.\ D {\bf 37}, 1237 (1988).

\bibitem{Zavattini:2007ee}
 E.~Zavattini {\it et al.},  
  Phys.\ Rev.\  D {\bf 77}, 032006 (2008)
  [arXiv:0706.3419 [hep-ex]].

\bibitem{Cameron:1993mr}
  R.~Cameron {\it et al.},
  Phys.\ Rev.\  D {\bf 47}, 3707 (1993).


\bibitem{battesti}
R.~Battesti {\it et al.},
  Eur.\ Phys.\ J. D {\bf 46}, 323 (2008).

\bibitem{Guendelman:2009kv}
  E.~I.~Guendelman, I.~Shilon, G.~Cantatore, K.~Zioutas,
  JCAP {\bf 1006}, 031 (2010).
  [arXiv:0906.2537 [hep-ph]].

\bibitem{Zavattini:2008cr}
  G.~Zavattini, E.~Calloni,
  Eur.\ Phys.\ J.\  {\bf C62}, 459-466 (2009).
  [arXiv:0812.0345 [physics.ins-det]].

\bibitem{Dobrich:2009kd}
  B.~Dobrich, H.~Gies,
  Europhys.\ Lett.\  {\bf 87}, 21002 (2009).
  [arXiv:0904.0216 [hep-ph]].

\bibitem{Jaeckel:2009dh}
  J.~Jaeckel,
  Phys.\ Rev.\ Lett.\  {\bf 103}, 080402 (2009)
  [arXiv:0904.1547 [hep-ph]].

\bibitem{Jaeckel:2010xx}
  J.~Jaeckel and S.~Roy,
  Phys.\ Rev.\  D {\bf 82}, 125020 (2010)
  [arXiv:1008.3536 [hep-ph]].

\bibitem{Tommasini:2009nh}
  D.~Tommasini, A.~Ferrando, H.~Michinel and M.~Seco,
  JHEP {\bf 0911}, 043 (2009)
  [arXiv:0909.4663 [hep-ph]].

\bibitem{Dobrich:2010hi}
  B.~Dobrich and H.~Gies,
  JHEP {\bf 1010}, 022 (2010)
  [arXiv:1006.5579 [hep-ph]].


\bibitem{Homma:2011pg}
  K.~Homma, D.~Habs, T.~Tajima,
  [arXiv:1103.1748 [physics.optics]].

\bibitem{Arik:2008mq}
  E.~Arik {\it et al.}  [CAST Collaboration],
  JCAP {\bf 0902}, 008 (2009)
  [arXiv:0810.4482 [hep-ex]].

\bibitem{Irastorza:2011gs} 
  I.~G.~Irastorza, F.~T.~Avignone, S.~Caspi, J.~M.~Carmona, T.~Dafni, M.~Davenport, A.~Dudarev and G.~Fanourakis {\it et al.},
  JCAP {\bf 1106}, 013 (2011)
  [arXiv:1103.5334 [hep-ex]].


\bibitem{Davidson:2000hf}
  S.~Davidson, S.~Hannestad and G.~Raffelt,
  JHEP {\bf 0005}, 003 (2000)
  [arXiv:hep-ph/0001179].

\bibitem{Ahlers:2009ru}
  M.~Ahlers, L.~A.~Anchordoqui, M.~C.~Gonzalez-Garcia,
  Phys.\ Rev.\  {\bf D81}, 085025 (2010).
  [arXiv:0910.5483 [hep-ph]].

\bibitem{Burrage:2009yz}
  C.~Burrage, J.~Jaeckel, J.~Redondo, A.~Ringwald,
  JCAP {\bf 0911}, 002 (2009).
  [arXiv:0909.0649 [astro-ph.CO]].


\bibitem{Masso:2006gc}
  E.~Masso and J.~Redondo,
  Phys.\ Rev.\ Lett.\  {\bf 97}, 151802 (2006)
  [arXiv:hep-ph/0606163].

\bibitem{Jaeckel:2006xm}
  J.~Jaeckel, E.~Masso, J.~Redondo, A.~Ringwald and F.~Takahashi,
  Phys.\ Rev.\  D {\bf 75}, 013004 (2007)
 [arXiv:hep-ph/0610203].

\bibitem{Melchiorri:2007sq}
  A.~Melchiorri, A.~Polosa and A.~Strumia,
  Phys.\ Lett.\  B {\bf 650}, 416 (2007)
  [arXiv:hep-ph/0703144].

\bibitem{Ehret:2009sq}
  K.~Ehret {\it et al.} [ALPS Collaboration ],
  Nucl.\ Instrum.\ Meth.\  {\bf A612}, 83-96 (2009).
  [arXiv:0905.4159 [physics.ins-det]].

\bibitem{Arias:2010bh}
  P.~Arias, J.~Jaeckel, J.~Redondo {\it et al.},
  Phys.\ Rev.\  {\bf D82}, 115018 (2010).
  [arXiv:1009.4875 [hep-ph]].

\bibitem{Hoogeveen:1992uk}
  F.~Hoogeveen,
  Phys.\ Lett.\  {\bf B288}, 195-200 (1992).

\bibitem{Hoogeveen:1990vq}
  F.~Hoogeveen, T.~Ziegenhagen,
  Nucl.\ Phys.\  {\bf B358}, 3-26 (1991).

\bibitem{Jaeckel:2007ch}
  J.~Jaeckel and A.~Ringwald,
  Phys.\ Lett.\  B {\bf 659}, 509 (2008)
  [arXiv:0707.2063 [hep-ph]].

\bibitem{Sikivie:2007qm}
  P.~Sikivie, D.~B.~Tanner and K.~van Bibber,
  Phys.\ Rev.\ Lett.\  {\bf 98}, 172002 (2007)
  [arXiv:hep-ph/0701198].

\bibitem{Rabadan:2005dm}
  R.~Rabadan, A.~Ringwald, K.~Sigurdson,
  Phys.\ Rev.\ Lett.\  {\bf 96}, 110407 (2006).
  [hep-ph/0511103].

\bibitem{Hartnett:2011zd} 
  J.~G.~Hartnett, J.~Jaeckel, R.~G.~Povey and M.~E.~Tobar,
  Phys.\ Lett.\ B {\bf 698}, 346 (2011)
  [arXiv:1101.4089 [quant-ph]].

\bibitem{Dobrich:2012sw} 
  B.~Dobrich, H.~Gies, N.~Neitz and F.~Karbstein,
  Phys.\ Rev.\ Lett.\  {\bf 109}, 131802 (2012)
  [arXiv:1203.2533 [hep-ph]].

\bibitem{Gies:2009wx}
  H.~Gies and J.~Jaeckel,
  JHEP {\bf 0908}, 063 (2009)
  [arXiv:0904.0609 [hep-ph]].

\bibitem{Heisenberg:1936qt}
W.~Heisenberg and H.~Euler,
Zeitschr.\ Phys.\ {\bf 98} 714  (1936). 

\bibitem{weisskopf}
V.~Weisskopf,
 Kong. Dans. Vid. Selsk. Math-fys. Medd. {\bf XIV}, 166
(1936). 
%

\bibitem{Schwinger:1951nm}
  J.~S.~Schwinger,
  Phys.\ Rev.\  {\bf 82}, 664 (1951).

\bibitem{Ahlers:2006iz}
  M.~Ahlers, H.~Gies, J.~Jaeckel and A.~Ringwald,
  Phys.\ Rev.\  D {\bf 75}, 035011 (2007)
  [arXiv:hep-ph/0612098].

\bibitem{King}
B.~King, A.~Di Piazza, and C.H.~Keitel, Nature Photon. {\bf 4}, 92 (2010). 

\bibitem{Tsai:1974ap}
  W.~y.~Tsai,
  Phys.\ Rev.\  D {\bf 10}, 2699 (1974).

\bibitem{Dittrich:2000zu}
  W.~Dittrich, H.~Gies,
  Springer Tracts Mod.\ Phys.\  {\bf 166}, 1-241 (2000).

\bibitem{Karbstein:2011ja} 
  F.~Karbstein, L.~Roessler, B.~Dobrich and H.~Gies,
  Int.\ J.\ Mod.\ Phys.\ Conf.\ Ser.\  {\bf 14}, 403 (2012)
  [arXiv:1111.5984 [hep-ph]].

\bibitem{BatShab}
I.~A.~Batalin and A.~E.~Shabad,
Sov.\ Phys.\ JETP {\bf 33}, 483 (1971).

\bibitem{Tsai:1974fa}
  W.~y.~Tsai and T.~Erber,
  {\it Phys.\ Rev.\  D}\ {\bf 10}, 492 (1974).

\bibitem{Tsai:1975iz}
  W.~y.~Tsai and T.~Erber,
  %
  Phys.\ Rev.\ D {\bf 12}, 1132, (1975).

\bibitem{Shabad:1975ik}
  A.~E.~Shabad,
  Annals Phys.\  {\bf 90}, 166 (1975).


\bibitem{Baier:2009it}
  V.~N.~Baier, V.~M.~Katkov,
  {\it Phys.\ Lett.\ A}\  {\bf 374}, 2201-2206 (2010)
  [arXiv:0912.5250 [hep-ph]].

\bibitem{DoebrichKarbstein}
B.~D\"obrich and F.~Karbstein, in preparation (2012).

\bibitem{Shabad:2003xy}
  A.~E.~Shabad,
  arXiv:hep-th/0307214.

\bibitem{shabad}
A.~E.~Shabad and V.~V.~Usov,
Astrophysics and Space Science {\bf 102}, 2 (1984).

\bibitem{Shabad:1972rg}
  A.~E.~Shabad,
  Lett.\ Nuovo Cim.\  {\bf 3S2}, 457 (1972).


\bibitem{Witte:1990}
N.~S.~ Witte,
J.\ Phys.A:Math.Gen. {\bf 23}, 5257 (1990).


\bibitem{Melrose:1976dr}
  D.~B.~Melrose and R.~J.~Stoneham,
  Nuovo Cim.\  A {\bf 32}, 435 (1976).

\bibitem{Tsai:1975tw}
  W.~-Y.~Tsai, T.~Erber,
  Acta Phys.\ Austriaca {\bf 45}, 245-254 (1976).
  
\bibitem{Cover:1974ij}
  R.~A.~Cover, G.~Kalman,
  Phys.\ Rev.\ Lett.\  {\bf 33}, 1113-1116 (1974).


\bibitem{Ahlers:2007rd}
  M.~Ahlers, H.~Gies, J.~Jaeckel, J.~Redondo and A. Ringwald,
  Phys.\ Rev.\  {\bf D76}, 115005 (2007).
  [arXiv:0706.2836 [hep-ph]].

\bibitem{Ahlers:2007qf}
  M.~Ahlers, H.~Gies, J.~Jaeckel, J.~Redondo and A. Ringwald,
  Phys.\ Rev.\  {\bf D77}, 095001 (2008).
  [arXiv:0711.4991 [hep-ph]].

\bibitem{Siegman}
{A.~E.~Siegman, \textit{Lasers}, First Edition, University Science Books, USA (1986).}

\bibitem{Baker:2011na} 
  O.~K.~Baker, M.~Betz, F.~Caspers, J.~Jaeckel, A.~Lindner, A.~Ringwald, Y.~Semertzidis and P.~Sikivie {\it et al.},
  arXiv:1110.2180 [physics.ins-det].

\bibitem{Dormicchi:1991ad} 
  O.~Dormicchi, R.~Penco, S.~Parodi, P.~Valente, A.~Bonito Oliva, G.~Gaggero, M.~Losasso and G.~Masullo {\it et al.},
  IEEE Trans.\ Magnetics {\bf 27}, 1958 (1991).

\bibitem{Meinke:1990yk} 
  R.~Meinke,
  IEEE Trans.\ Magnetics {\bf 27}, 1728 (1991).

\bibitem{Gies:2006hv} 
  H.~Gies, J.~Jaeckel and A.~Ringwald,
  Europhys.\ Lett.\  {\bf 76}, 794 (2006)
  [hep-ph/0608238].

\bibitem{Knabbe}
Ernst-Axel Knabbe, private communication.

\bibitem{Schubert:2000yt}
  C.~Schubert,
  Nucl.\ Phys.\  B {\bf 585}, 407 (2000)
  [arXiv:hep-ph/0001288].

\bibitem{Gradshteyn}
I.~S.~Gradshteyn and I.~M.~Ryzhik, \textit{Table of Integrals, Series, and Products}, Fifth Edition, Academic Press, UK (1994).

\end{thebibliography}
\end{document}